\documentclass[aps, prl, reprint, superscriptaddress]{revtex4-1}

\pdfoutput=1 
\usepackage{graphicx}
\usepackage{color,amsmath}
\usepackage[colorlinks,linkcolor=blue,citecolor=blue,urlcolor=blue]{hyperref}
\usepackage{siunitx}
\usepackage{enumitem}

\begin{document}

\title{Electron-Hole Interference in an Inverted-Band Semiconductor Bilayer}

\author{Matija Karalic}
\thanks{These authors contributed equally to this work.}
\affiliation{Solid State Physics Laboratory, ETH Zurich, 8093 Zurich, Switzerland}

\author{Antonio \v{S}trkalj}
\thanks{These authors contributed equally to this work.}
\affiliation{Institute for Theoretical Physics, ETH Zurich, 8093 Zurich, Switzerland}

\author{Michele Masseroni}
\affiliation{Solid State Physics Laboratory, ETH Zurich, 8093 Zurich, Switzerland}

\author{Wei Chen}
\affiliation{Institute for Theoretical Physics, ETH Zurich, 8093 Zurich, Switzerland}
\affiliation{National Laboratory of Solid State Microstructures and Department of Physics, Nanjing University, Nanjing 210093, China}

\author{Christopher Mittag}
\affiliation{Solid State Physics Laboratory, ETH Zurich, 8093 Zurich, Switzerland}

\author{Thomas Tschirky}
\affiliation{Solid State Physics Laboratory, ETH Zurich, 8093 Zurich, Switzerland}

\author{Werner Wegscheider}
\affiliation{Solid State Physics Laboratory, ETH Zurich, 8093 Zurich, Switzerland}

\author{Thomas Ihn}
\affiliation{Solid State Physics Laboratory, ETH Zurich, 8093 Zurich, Switzerland}

\author{Klaus Ensslin}
\affiliation{Solid State Physics Laboratory,  ETH Zurich, 8093 Zurich, Switzerland}

\author{Oded Zilberberg}
\affiliation{Institute for Theoretical Physics, ETH Zurich, 8093 Zurich, Switzerland}

\date{\today}

\begin{abstract}
Electron optics in the solid state promises new functionality in electronics through the possibility of realizing micrometer-sized interferometers, lenses, collimators and beam splitters that manipulate electrons instead of light. Until now, however, such functionality has been demonstrated exclusively in one-dimensional devices, such as in nanotubes, and in graphene-based devices operating with \textit{p-n} junctions. In this work, we describe a novel mechanism for realizing  electron  optics  in  two  dimensions. By studying a two-dimensional Fabry--P\'{e}rot interferometer based on a resonant cavity formed in an InAs/GaSb double quantum well using \textit{p-n} junctions, we establish that  electron-hole hybridization in band-inverted systems can facilitate  coherent interference.  With this discovery, we expand the field of electron optics to encompass materials that exhibit band inversion and hybridization, with the promise to surpass the performance of current state-of-the-art devices.
\end{abstract}

\maketitle

Common interferometers in optics, such as the Fabry--P\'{e}rot or the Mach--Zehnder interferometer, rely on the interference of monochromatic waves with the same propagation direction. In this sense they can be regarded as one-dimensional. In an optical  Fabry--P\'{e}rot interferometer, the interference pattern can be observed as a periodic change of maxima and minima in the  transmitted intensity while the wavelength of the light is gradually varied. However, the wavelength period of the interference pattern depends on the incident angle of the light because it is the projection of the wave vector onto the interferometer axis that enters the conditions for constructive and destructive interference. As a result, the interference pattern is averaged out  if light of all possible incident angles is sent through the interferometer at once. The same arguments apply to electronic Fabry--P\'{e}rot interferometers. Pronounced interference is usually observed only in one-dimensional systems, such as carbon nanotubes \cite{liang_fabry--perot_2001} or quantum Hall edge states \cite{van_wees_observation_1989,ji_electronic_2003}, where the propagation direction of electrons is restricted to one dimension.

Recently, electronic interference has been discovered in two-dimensional $\textit{pnp-}$ (or $\textit{npn-}$) junctions of single- and multilayer graphene \cite{young_quantum_2009,campos_quantum_2012,varlet_fabry-perot_2014}, where electrons (or holes) are incident under all possible angles. There, Klein or anti-Klein tunneling in conjunction with  Dirac's spectrum lead to a specific selectivity for incident angles that enables the observation of interference in these two-dimensional geometries without the need for additional confinement in the transverse direction. Together with the realization of electron lenses, collimators and beam splitters \cite{spector_electron_1990,cheianov_focusing_2007,lee2015,park_electron_2008,rickhaus_gate_2015, chen_electron_2016}, this discovery has marked a leap forward for electron optics.

Here, we report an interference mechanism fundamentally different from that in graphene by studying transport through a two-dimensional Fabry--P\'{e}rot interferometer realized using $\textit{p-n}$ junctions in an inverted InAs/GaSb double quantum well. The interference is facilitated by electron-to-hole scattering in the band-inverted regime. Such a scenario effectively produces parallel one-dimensional transport channels which share almost identical conditions for constructive interference, leading to a nonvanishing interference pattern even after angle averaging. 
With this finding, we expand the field of electron optics to include materials that exhibit band inversion and hybridization. 

\begin{figure*}
	\includegraphics[]{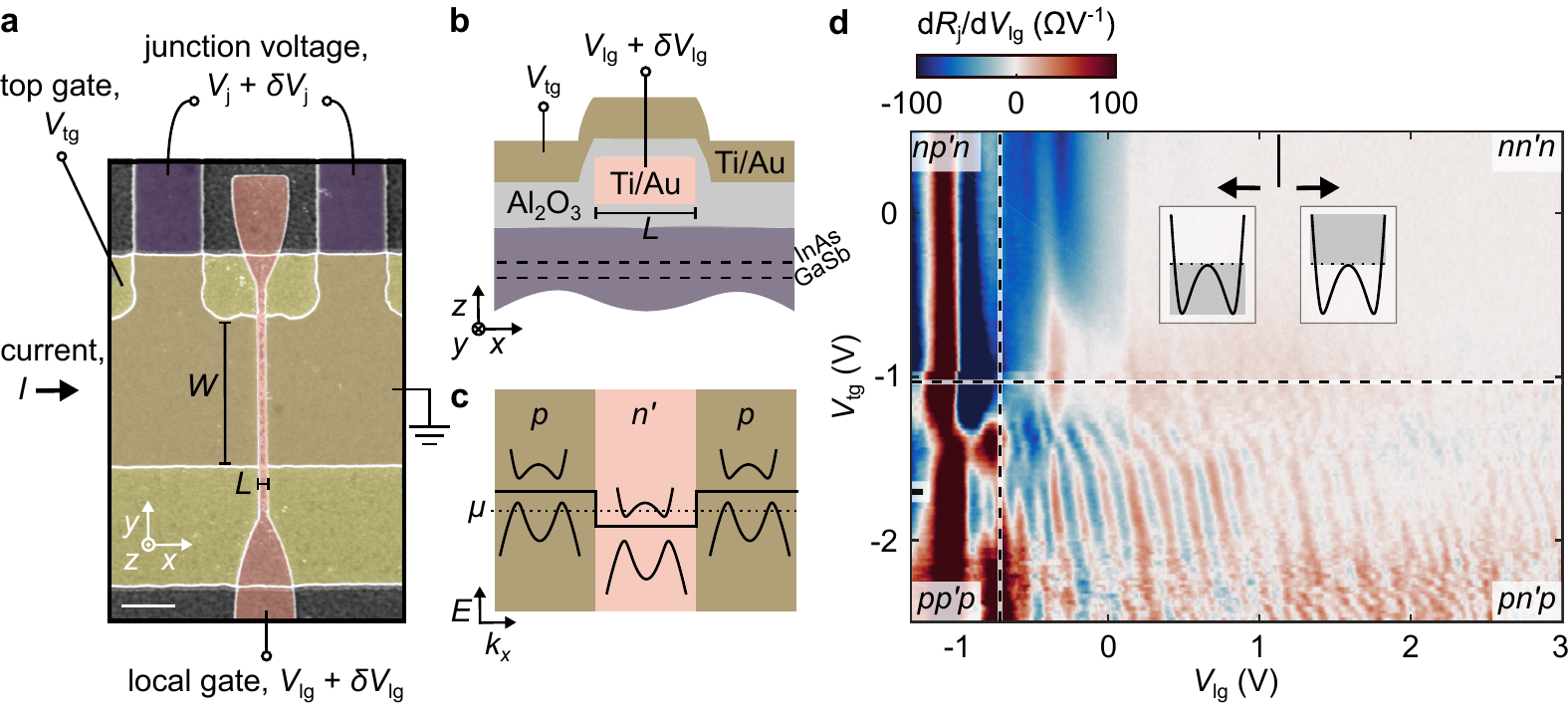}
	\caption{{\bf Device structure and phase diagram.} \textbf{a}, False-color image describing the device structure and operation. The scale bar represents $\SI{2}{\micro\meter}$. \textbf{b}, Schematic device cross-section. \textbf{c}, Exemplary band alignment $E(k)$ in the $pn'p$ configuration; $\mu$ is the electrochemical potential, indicated by the dotted line, and the solid line indicates the charge neutrality point (CNP) in each region. \textbf{d}, Color map depicting the differential junction resistance $\mathrm{d}R_\mathrm{j}/\mathrm{d}V_\mathrm{lg}$ as a function of $V_\mathrm{lg}$ and $V_\mathrm{tg}$ at $1.3$\,K. The resulting phase diagram is subdivided into quadrants according to the charge carrier configuration. The horizontal and vertical dashed lines mark the CNPs of the leads and the cavity, respectively. The insets show the parts of the band structure probed to the left and right of the solid line at $V_\mathrm{lg} = 1.1$\,V in the $pn'p$ configuration, as highlighted by the shading.} 
	\label{fig1}
\end{figure*}

Figure 1a is a false-color scanning electron micrograph showing a zoom-in of a typical device used in this work. The device is defined by a combination of overlapping metallic gates, see Fig.~1b, placed on top of a semiconductor heterostructure hosting the coupled InAs and GaSb quantum wells.  Charge carriers are confined in two dimensions to the planes of the buried quantum wells, which  
extend in $x\textit{-}y$ direction and have thicknesses in $z$ of $13.5$ (InAs) and $8$\,nm (GaSb), respectively. A local gate of length $L$ spanning the full device width is electrically separated from the heterostructure surface below and the top gate above by an insulating oxide layer. Below the local gate, the density of charge carriers in the quantum wells is affected by this gate only. The adjoining outer regions (leads) are affected by the top gate exclusively (see Supplementary Information Fig.~S1 for a full electrostatic simulation). A typical device contains multiple local gates of varying lengths, allowing us to study the length dependence of the interference in the same device. 
We inject a dc current $I \sim$ 100 nA  at $1.3$\,K and record the resulting junction voltage $V_\mathrm{j}$ as a function of the voltage applied to the top gate, $V_\mathrm{tg}$, and the voltage applied to the local gate, $V_\mathrm{lg}$ (Figs.\ 1a and b). In order to probe the locally gated region directly, we also apply a small ac modulation to the local gate, $\delta V_\mathrm{lg}$, and pick up the corresponding ac response in the junction voltage,  $\delta V_\mathrm{j}$. Then, $\delta V_\mathrm{j}/\delta V_\mathrm{lg} I$ represents the quantity $\mathrm{d}R_\mathrm{j}/\mathrm{d}V_\mathrm{lg}$ and is sensitive solely to changes in the transmission of the locally gated region. From now on, for reasons to become apparent shortly, we will refer to the locally gated region as the ``cavity''.   

In the InAs/GaSb bilayer system, which has recently been studied due to its topological insulator properties \cite{liu_quantum_2008,suzuki_edge_2013, du_robust_2015,pribiag_edge-mode_2015}, electron-like and hole-like states coexist in the energetic interval characterized by band inversion~\cite{altarelli_electronic_1983,lakrimi_minigaps_1997,yang_evidence_1997,cooper_resistance_1998}, within which the charge neutrality point (CNP) resides. To explore the available parameter space, we tune the voltages $V_\mathrm{tg}$ and $V_\mathrm{lg}$ to obtain a variety of charge carrier configurations in the device. 
We label the configurations with $(x, y',x)$, where $x \in \{n, p\}$ denotes the density of majority charge carriers in the leads, and $y' \in \{n', p'\}$ that of majority charge carriers in the cavity, where $n$, $n'$ and $p$, $p'$ are the densities of electron-like and hole-like charge carriers, respectively. 
In a $pn'p$ configuration, there can also be minority holes in the cavity ($p'$), and minority electrons in the leads ($n$). Figure 1c depicts the corresponding simplified band lineup. 

The measurement in Fig.~1d, which contains the main findings of this work, displays $\mathrm{d}R_\mathrm{j}/\mathrm{d}V_\mathrm{lg}$ as a function of $V_\mathrm{lg}$ and $V_\mathrm{tg}$ for a local gate of lithographic length $L = 320$\,nm. The quadrants representing the possible charge carrier configurations indicated in the figure are separated by the experimentally determined CNPs in the leads and the cavity (horizontal and vertical dashed lines). We recognize regular oscillations in the $pn'p$ configuration (nearly vertical stripes of alternating minima and maxima) which depend strongly on $V_\mathrm{lg}$, and weakly on $V_\mathrm{tg}$. In the other quadrants, no such oscillations appear. The observed oscillations persist up to $V_\mathrm{lg} \sim 1$\,V before disappearing. Concomitantly, we estimate from independent density measurements (Supplementary Information Fig.\ S2) that $p'$ vanishes at $V_\mathrm{lg} \approx 1.1$\,V as the Fermi energy crosses the top edge of the ground state subband in the unhybridized valence band (see insets of Fig.~1d). 
In other words, when $V_\mathrm{lg} > 1.1$\,V we obtain a conventional electron system devoid of holes. This observation suggests that the coexistence of electrons and holes in the cavity is crucial for the emergence of the resistance oscillations.

\begin{figure*}
	\includegraphics[]{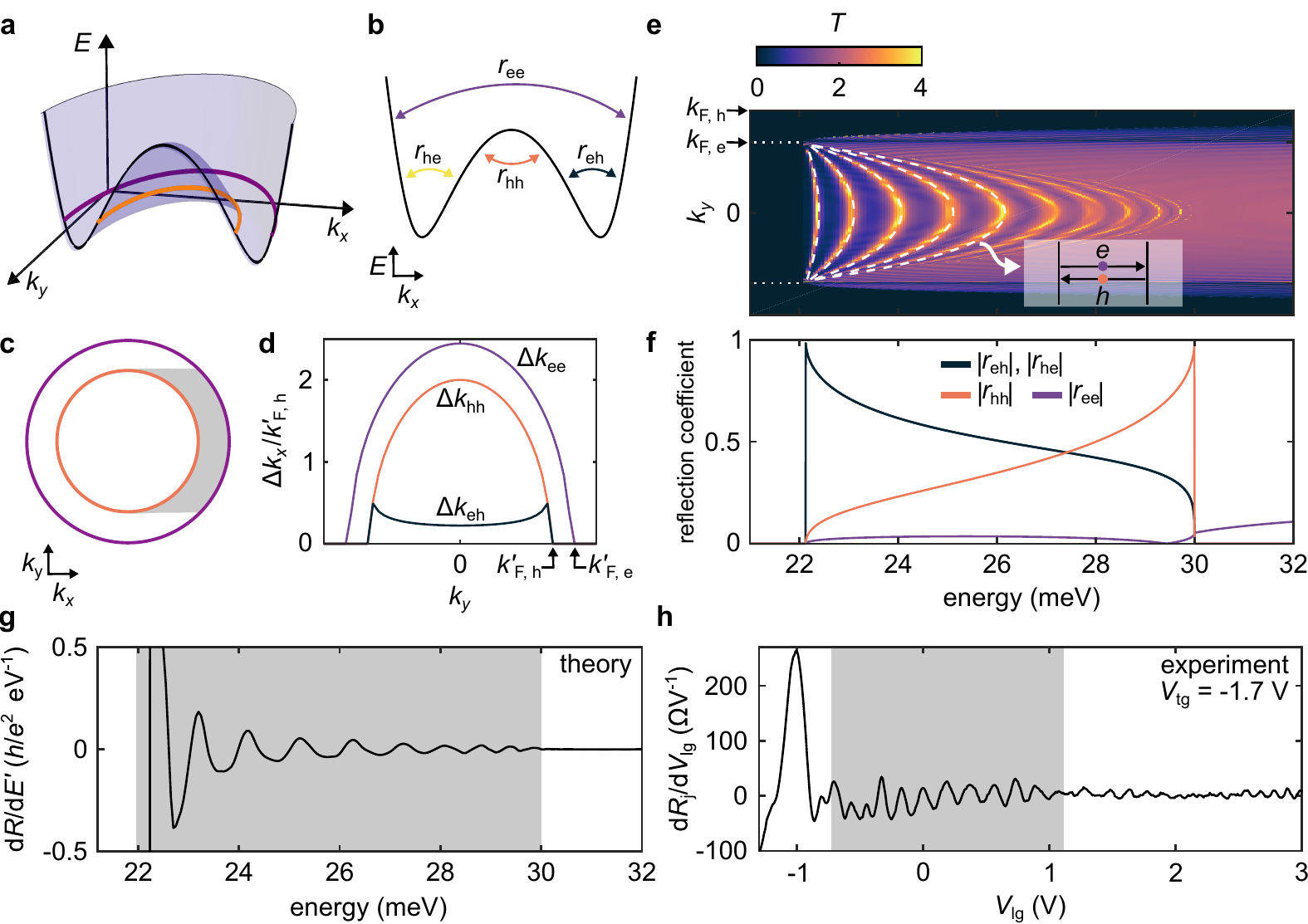}
	\caption{{\bf Theoretical modelling of the interference.}  \textbf{a}, Sombrero hat dispersion $E(k_x, k_y)$ of the hybridized conduction band resulting from the theoretical model, equation~\eqref{hamiltonian}, as applied to the InAs/GaSb system.
		\textbf{b}, Overview of the scattering processes $ij$ that occur within the $n'$-doped cavity when a quasiparticle impinges onto the interface between cavity  and lead. The $r_{ij}$ are reflection coefficients.
		\textbf{c}, Two Fermi circles from \textbf{a}, with the larger (smaller) circle being the Fermi circle of electron-like (hole-like) states. The shaded region marks the momentum transfer $\Delta k_x$ for the $eh$ scattering processes assuming conservation of $k_y$; $\Delta k_x$ does not depend strongly on $k_y$.
		\textbf{d}, Momentum transfer $\Delta k_x$ as a function of $k_y$ for the various scattering processes $ij$ from \textbf{b}. Here, $\Delta k_\mathrm{eh}$ and $\Delta k_\mathrm{he}$ are identical. $k_\mathrm{F,\,e}$ ($k'_\mathrm{F,\,e}$) and $k_\mathrm{F,\,h}$ ($k'_\mathrm{F,\,h}$) are Fermi wave numbers of electron-like and hole-like states in the leads (cavity), respectively.
		\textbf{e}, Transmission $T$ through the $pn'p$ junction as a function of the Fermi energy $E'$ in the $n'$-doped cavity and the momentum $k_y$ of the incident wave, calculated using a scattering matrix approach. The dashed lines denote the constructive interference condition assuming $eh$ and $he$ scattering only (equation~\eqref{simplification} and inset). The maximum possible value of $T$ is 4 due to having two orbital and two spin degrees of freedom.
		\textbf{f}, Reflection amplitudes $\lvert r_{ij} \rvert$ for the various scattering processes $ij$ from \textbf{b}, plotted as a function of $E'$ and for $k_y=0$.
		\textbf{g}, Differential resistivity $\mathrm{d}R/\mathrm{d}E'$ as a function of $E'$ obtained from \textbf{e} using equation ~\eqref{conductance}. Oscillations are present in the shaded energy region in which electrons and holes coexist.
		\textbf{h}, Cut through Fig.~1d at $V_\mathrm{tg} = -1.7$\,V, with the shaded region above the CNP marking electron and hole coexistence.} 
	\label{fig2}
\end{figure*}

In the $pn'p$ configuration, \textit{p-n} junctions delimit the cavity. Across each junction, the density of states exhibits a local minimum as the Fermi energy transitions smoothly from below to above the hybridization gap~\cite{karalic_lateral_2017}. Specifically, if a true gap exists, particles must tunnel in order to be transmitted.  The \textit{p-n} junctions can therefore act as semitransparent mirrors, trapping particles in the cavity and enabling the formation of discrete standing-wave modes. The transmission, and accordingly, the resistance, of the resulting resonant cavity is then modulated as a function of the density $n'$ within it, with constructive and destructive interference alternating in a periodic manner. In optics, this type of interferometer is known as a Fabry--P\'{e}rot etalon \cite{saleh_fundamentals_2019}.

To better understand the nature of the resistance oscillations, we systematically investigate $\textit{pn'p}$ junctions with local gates of various lengths, both in a single device and across multiple devices, and find the oscillations to be reproducible (Supplementary Information Figs.\ S3, S4, S9, S10). This robustness also holds for multiple cooldowns, with the pattern of minima and maxima hardly changing despite repeated thermal cycling. We observe that both the oscillation amplitude and the average period in $V_\mathrm{lg}$ decrease with increasing $L$ (Supplementary Information Fig.\ S4). Furthermore, the oscillations remain unchanged if the current direction is reversed, and their amplitude also decreases with increasing current bias (Supplementary Information Fig.\ S5). A similar decrease occurs at elevated temperatures, with the oscillations completely disappearing above $\sim 4$\,K (Supplementary Information Fig.\ S6). 

Our experimental results are consistent with the emergence of standing wave modes in the cavity, such that the oscillations arise due to quantum interference in a two-dimensional electronic analogue of the Fabry--P\'{e}rot etalon. The decrease in amplitude with increasing $L$ is in agreement with a finite phase coherence length. From the temperature dependence, we extract an energy scale of $\sim 1$\,meV for the mean level spacing in the cavity (Supplementary Information Fig.\ S6), which is in agreement with theoretical calculations, as discussed below.

\begin{figure*}
	\includegraphics[]{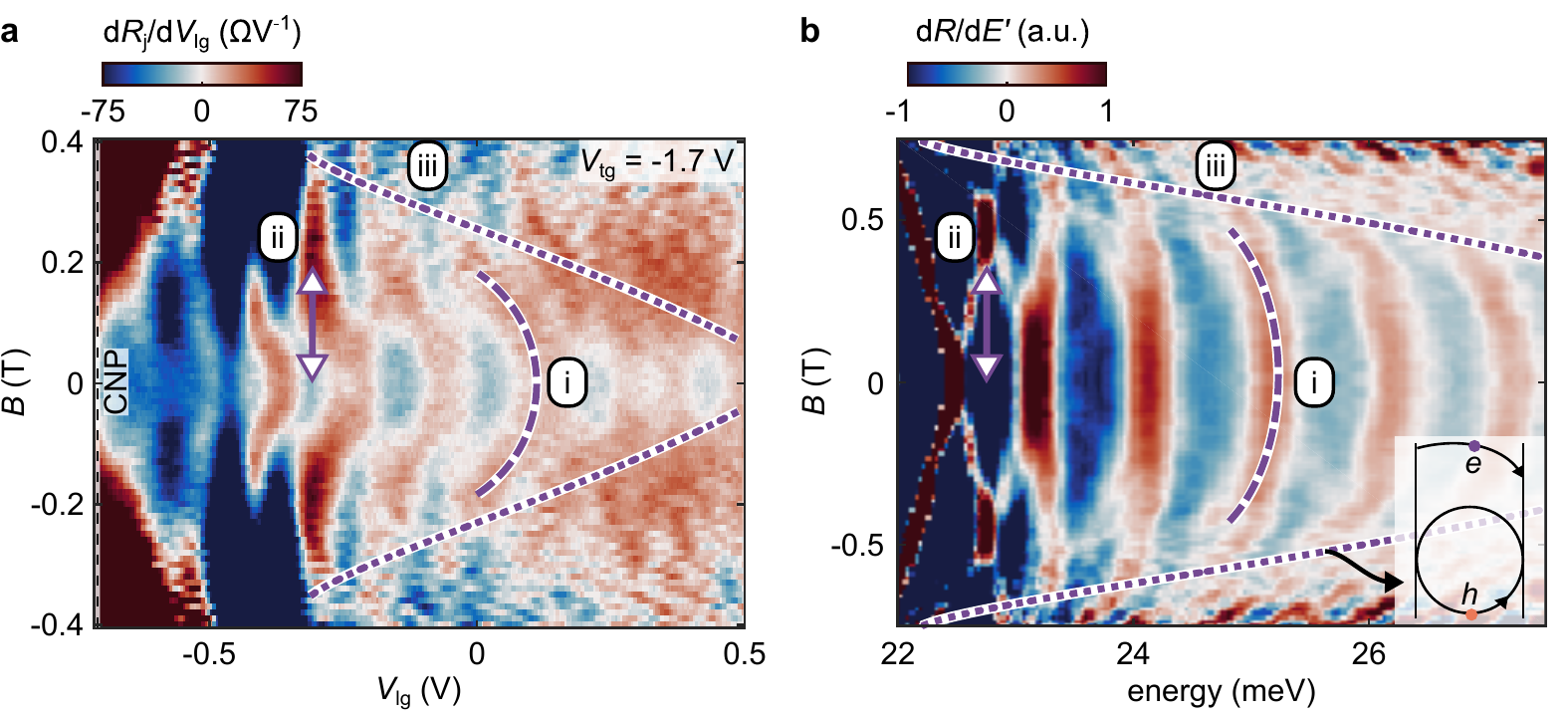}
	\caption{{\bf Magnetic field dependence.} \textbf{a}, Color map depicting the differential junction resistance $\mathrm{d}R_\mathrm{j}/\mathrm{d}V_\mathrm{lg}$ as a function of $V_\mathrm{lg}$ and $B$ at $V_\mathrm{tg} = -1.7$\,V  and  $1.3$\,K. \textbf{b}, Calculated dependence of $\mathrm{d}R/\mathrm{d}E'$ on the Fermi energy $E'$ in the cavity and $B$. The inset shows schematic trajectories for electron-like and hole-like states at the critical magnetic field $B_\mathrm{c}$ of the hole-like states,  represented in both \textbf{a} and \textbf{b} by the dotted lines. Dashed lines and arrows indicate features common to both experiment and calculation, discussed under points (i)--(iii) in the main text.} 
	\label{fig3}
\end{figure*}

In order to understand the origin of the interference, we model the hybridized electron-hole system using the minimal Bernevig-Hughes-Zhang (BHZ) Hamiltonian \cite{bernevig_quantum_2006-1}
\begin{equation}
H= 	
\begin{pmatrix}
h(\mathbf{k}) & 0 \\
0 & h^*(-\mathbf{k})
\end{pmatrix},
\label{hamiltonian}
\end{equation}
where $h(\mathbf{k}) = \mathcal{B} \mathbf{k}^2 \hat{\mathbf{1}} + (\mathcal{M}_0 + \mathcal{M}_2 \mathbf{k}^2) \hat{\sigma}_z + \mathcal{A} k_x \hat{\sigma}_x - \mathcal{A} k_y \hat{\sigma}_y$, $\hat{\sigma}_i$ are Pauli matrices operating on the orbital degrees of freedom and $\mathbf{k}=(k_x, k_y)$ is the wave vector. The non-zero blocks in the Hamiltonian act on the two spin degrees of freedom separately. The parameter $\mathcal{M}_0$ governs the band inversion, $\mathcal{M}_2$ symmetrically controls the band curvatures (effective masses), $\mathcal{B}$ is a symmetry breaking term between electron and hole bands and $\mathcal{A}$ determines the electron-hole coupling.
In Fig.~2a, we present the dispersion of the hybridized conduction band relevant for the cavity in the $pn'p$ configuration. In the band-inverted regime, the bulk band structure has the shape of a ``Sombrero  hat'', formed due to the hybridization between the InAs conduction and GaSb valence bands.

We assume that both cavity and leads are infinite in $y$-direction, such that $k_y$ remains a good quantum number. Scattering between the $n'$-doped cavity and $p$-doped leads conserves $k_y$, such that $k_y=k_y^\prime$ (primed quantities refer to the cavity, unprimed ones to the leads). We therefore treat the two-dimensional scattering problem as an infinite set of one-dimensional wires with different $k_y$ momenta. For each $k_y$, we note that there are four possible scattering processes of quasiparticles that can occur at the interfaces between the cavity and the leads (Fig.~2b),  distinguishing scattering between (i) alike particles---electron-to-electron and hole-to-hole ($ee$ and $hh$)---and (ii) different particles---electron-to-hole and vice versa ($eh$ and $he$). This situation is reminiscent of optical Fabry--P\'{e}rot cavities with birefringent (polarization-dependent) mirrors~\cite{okada_electronic_1975}. Crucial to our result is that the wave vector change $\Delta k_x$ for scattering processes $eh$ and $he$ does not depend strongly on $k_y$ at energies close to the gap (Figs.~2c and 2d). 
We solve the scattering problem analytically and obtain the transmission $T$ for each $k_y$ wire as a function of the Fermi energy $E'$ in the cavity (Fig.~2e and Methods section). 
The maxima of the transmission depend only weakly on $k_y$ for energies close to the band bottom, near the hybridization gap. Away from the band bottom, the dependence on $k_y$ increases and transmission maxima curve towards lower energies. 

Upon inspection of the reflection amplitudes $|r_{ij}|$ at the interface between the cavity and the leads (see Fig.~2b), we identify that the $eh$ and $he$ scattering processes dominate the transmission $T$ at energies close to the gap (Fig.~2f). 
With this in mind, we approximate the condition for constructive interference as
\begin{equation}
\Delta k_\mathrm{eh / he} L = 2\pi m,
\label{simplification}
\end{equation}
where $\Delta k_\mathrm{eh/he}$ is the energy-dependent momentum transfer in $x$-direction in the $eh$ and $he$ scattering processes and $m \in N$ is the ordinal number of the transmission maxima. This approximation agrees well with the calculated $T$, see Fig.~2e. 

To obtain the conductance, we integrate the calculated $T$ over all incident states with different momenta $k_y$,
\begin{equation}
G(E') = \frac{2 e^2}{h} \frac{W}{2\pi} \int_{-k_\mathrm{F}}^{k_\mathrm{F}} \mathrm{d} k_y \, T(E', k_y),
\label{conductance}
\end{equation}
and set the resistance $R(E') = 1/G(E')$. The width of the sample in $y$-direction is $W$ and the integration boundaries, $\pm k_\mathrm{F}$, are given by the Fermi wave number associated with the larger of the two Fermi circles in the leads.  
The numerical derivative $\mathrm{d}R/\mathrm{d}E'$ is shown in Fig.~2g. Oscillations appear when the Fermi energy in the cavity lies in the energy region where the coherent $eh$ and $he$ processes exist, indicated by the shaded area in Fig.~2g. Near the gap, the oscillation period grows with increasing energy. In contrast, the oscillation amplitude decays with increasing energy as the $ee$ and $hh$ scattering processes become relevant. Once the holes in the cavity are depleted and the Fermi energy enters the region above the top of the Sombrero hat, the oscillations disappear. We postulate that this behavior is a generic property of band structures with the shape of a Sombrero hat. The interference mechanism is a result of both the selective scattering of electrons-to-holes and the fact that the interference condition does not vary much with $k_y$.

We now analyze the periodicity of the experimentally observed oscillations (Fig.~2h) with the theoretical insights in mind. The aforementioned features, namely the increasing period and the decreasing amplitude, both as a function of increasing energy, are present in the experiment.
Furthermore, by converting gate voltage to density, and then relating the density to the wave number in the cavity, we obtain an estimate for the cavity length based on the period (Supplementary Information Fig.\ S7). The estimate agrees with the lithographic gate length.   
It is worth emphasizing that in the case of dominant $ee$ scattering, the oscillations would grow in amplitude with increasing electron density, or equivalently with increasing voltage. The opposite is true in the experiment, providing confirmation of the significance of electron-hole interference.

In the $\textit{np'n}$ configuration no oscillatory features appear, even upon extension of the voltage range to more negative $V_\mathrm{lg}$ (Supplementary Information Fig.\ S9). The origin of the lack of oscillations in this regime is currently an open question. Note that the band structure of our system lacks particle-hole symmetry.

We now turn to the dependence of the resistance oscillations on a perpendicular magnetic field $B$. In the experiment (Fig.~3a), we track the evolution of the oscillations close to the CNP as a function of $B$. 
Overall, lines of constant transmission (phase) are initially independent of $B$ close to $B = 0$. They then bend towards lower $V_\mathrm{lg}$ with increasing $B$. This bending ceases thereafter as the lines straighten once again before the resistance oscillations vanish. Comparing the $\mathrm{d}R_\mathrm{j}/\mathrm{d}V_\mathrm{lg}$ signal at $B = 0$ and at $B = 0.25$\,T, we recognize that minima and maxima are interchanged. Such a shift of half a period is equivalent to a phase shift of $\pi$.

Placing model \eqref{hamiltonian} on a lattice \cite{groth_kwant:_2014}, we numerically calculate the expected magnetic field dependence, see Fig.~3b. 
Crucially, opposite spin states possess different responses to the magnetic field. The resulting plot bears similarities with the experimental measurement, and qualitatively captures all striking observations, namely (i) the qualitative bending of the lines, (ii) the exchange of minima and maxima and (iii) the energy-dependent (gate-voltage dependent) critical magnetic field above which oscillations are no longer present.
From the model analysis, we conclude that (i) highlights the hole-like nature of the cavity states, (ii) is related to the spin splitting of the initially degenerate standing wave modes in the cavity and (iii) is a result of the decreasing density of hole-like states (see below).
Note that the minimal model \eqref{hamiltonian} excludes several complications present in the experiment, such as band anisotropy, additional spin--orbit coupling terms, a density-dependent band overlap and the fact that the electrostatics of the device change as a function of the gate voltages. These effects may conspire to produce the quantitative differences between the two plots in Fig.~3.

Semiclassically, we expect the interference phenomenon to be suppressed with increasing magnetic field as the cyclotron radius $r_\mathrm{c} = \hbar k_\mathrm{F}'/eB$ decreases below a threshold related to the length $L$ of the cavity, where $k_\mathrm{F}'$ is the Fermi wave number in the cavity. Namely, when $2 r_\mathrm{c} = L$ at a critical field $B_\mathrm{c}$, quasiparticles cannot traverse the cavity any longer. As the hole-like states have smaller $k_\mathrm{F}'$ than the electron-like states, their $B_\mathrm{c}$ occurs earlier, see dotted lines and the inset in Fig.~3b. While it is not straightforward to determine $B_\mathrm{c}$ in the experiment (the corresponding lines in Fig.~3a are guides to the eye only), in the model analysis the oscillations essentially disappear when this condition is fulfilled for the hole-like states. Therefore, we conclude that in the $pn'p$ configuration the minority holes in the cavity decisively govern the interference dynamics in the presence of a magnetic field and not the majority electrons.

Our results set the stage for engineering electron optics phenomena in a variety of materials that feature band inversion and hybridization, or equivalently, a Sombrero hat dispersion. Examples include systems with strong Rashba spin--orbit interaction~\cite{bychkov_properties_1984}, two-dimensional monochalcogenides like GaSe~\cite{li_controlled_2014}, transition metal dichalcogenides like WTe$_2$~\cite{qian_quantum_2014} and three-dimensional topological insulators, such as those based on bismuth~\cite{zhang_topological_2009}. Furthermore, we predict that single \textit{p-n} junctions in these materials will exhibit focusing of electrons and holes due to electron-to-hole scattering, thus enabling the implementation of a range of devices such as Veselago lenses~\cite{cheianov_focusing_2007,lee2015} and beam splitters~\cite{rickhaus_gate_2015}. 
In the particular case of InAs/GaSb, the electrostatic tunability with back and front gates is advantageous for precise control of the band structure~\cite{qu_electric_2015} and hence the interference phenomenon. Compatibility with standard large-scale semiconductor processing techniques enables the straightforward realization of networks of interferometers, which is favorable for upscaling.

\begin{acknowledgments}
We thank P.\ Rickhaus for helpful discussions. This work was supported by the ETH FIRST laboratory. We acknowledge financial support from the Swiss National Science Foundation through the National Center of Competence in Research on Quantum Science and Technology (NCCR QSIT) and through grant PP00P2 163818. 

M.K.\ conceived the experiment with input from K.E\ and T.I.; T.T.\ and W.W.\ provided the heterostructures used in this work, from which M.K.\ and C.M.\ fabricated devices; M.K.\ and M.M.\ performed the measurements and analyzed them; A.\v{S}., W.C., and O.Z.\ analyzed the theoretical model describing the experiment; M.K., A.\v{S}., T.I., K.E.\ and O.Z.\ wrote the manuscript with input from all authors. 

Correspondence and requests for materials should be addressed to M.K.\ (\href{mailto:makarali@phys.ethz.ch}{makarali@phys.ethz.ch}), W.C.\ (\href{mailto:pchenweis@gmail.com}{pchenweis@gmail.com}) or T.I.\ (\href{mailto:ihn@phys.ethz.ch}{ihn@phys.ethz.ch}).
\end{acknowledgments}



%

\widetext 

\newpage
\cleardoublepage
\renewcommand{\figurename}{Extended Data Figure}

\onecolumngrid
\begin{center}
	\textbf{\normalsize Supplemental Material for}\\
	\vspace{3mm}
	\textbf{\large Electron-Hole Interference in an Inverted-Band Semiconductor Bilayer}\\
	\vspace{4mm}
	{ Matija Karalic${}^{1,*}$, Antonio \v{S}trkalj${}^{2,*}$, Michele Masseroni${}^1$, Wei Chen${}^{2,3}$, Christopher Mittag${}^1$, Thomas Tschirky${}^1$, Werner Wegscheider${}^1$, Thomas Ihn${}^1$, Klaus Ensslin${}^1$, and Oded Zilberberg${}^2$}\\
	\vspace{1mm}
	\textit{\small ${}^1$ Solid State Physics Laboratory, ETH Zurich, 8093 Zurich, Switzerland}\\
	\textit{\small ${}^2$ Institute for Theoretical Physics, ETH Z\"urich, 8093 Z\"urich, Switzerland}\\
	\textit{\small ${}^3$ National Laboratory of Solid State Microstructures and Department of Physics, Nanjing University, Nanjing 210093, China}
	
	\vspace{5mm}
*These authors contributed equally to this work.
\end{center}

\section{Device specifications}
\subsection{Device fabrication} 
Our InAs/GaSb double quantum wells (QWs) are embedded in a semiconductor heterostrucutre grown using molecular-beam epitaxy on a Te $n$-doped GaSb substrate, as illustrated in Extended Data Fig.\ 1. The crystallographic growth direction is [100]. The substrate, meant to serve as a back gate, is electrically isolated from the active layers by a buffer consisting of a GaSb layer and a GaSb/AlGaSb superlattice. The QWs are straddled by Al(Ga)Sb confinement layers, and  a GaSb cap protects the underlying layers from reacting with the environment. The heterostructure is nominally undoped, however due to the unintentional doping present in InAs and AlSb containing heterostructures, the QWs are intrinsically $n$-doped, meaning that there is an excess of electrons, $n > p$, in the absence of any applied gate voltages.

To fabricate devices from the heterostructures, we first define the mesa by wet chemical etching. Then, we deposit the first layer of insulating Al$_2$O$_3$, followed by evaporation of the local gates, typically three per device and having different length $L$. We subsequently deposit another layer of Al$_2$O$_3$, and then evaporate the global top gate. The final step is the electrical contacting of the QWs and the buried local gates. This is achieved by etching the oxide stack and the heterostructure down to the InAs QW, the latter required for the ohmic contacts only, and depositing metal. Extended Data Fig. 2 shows a scanning electron micrograph of a completed device. The exact fabrication procedure is as follows:
\begin{enumerate}[noitemsep,wide, labelwidth=!, labelindent=0pt]
	\item Preparation:
	\begin{itemize}
		\item Cleave a $3 \times 2.8$\,mm$^2$ piece from the wafer using a diamond wafer scriber.
		\item Clean with warm acetone and IPA ($\SI{50}{\celsius}$), applying gentle sonication if necessary.
	\end{itemize}
	\item Mesa structure definition:
	\begin{itemize}
		\item Perform dehydration bake for 120\,s at $\SI{120}{\celsius}$.
		\item Spin photoresist AZ 1505 at 3000 rpm for 3\,s by accelerating for 3\,s from standstill, then accelerating for 5\,s to 5000 rpm and holding for 60\,s. 
		\item Perform prebake for 120\,s at $\SI{120}{\celsius}$. The resulting photoresist thickness is around 500\,nm. 
		\item Remove edge beads from the sample by exposing the edges using contact printing with a square pattern on a quartz mask in a Karl S\"{u}ss MJB3 mask aligner (400\,nm 350\,W Hg light source). The exposure lasts 10\,s, and is followed by developing in Microposit MF319 for 15\,s and rinsing in DI water for 20\,s.
		\item Perform exposure of the structure proper for 3.2\,s similar to above, then develop in Microposit MF319 for 20\,s and rinse in DI water for 20\,s.
		\item Perform optional postbake for 60\,s at $\SI{120}{\celsius}$.
		\item Etch the heterostructure to define the mesa. The etching solution is a citric acid based etchant consisting of DI water, C$_6$H$_8$O$_7$ (50\% diluted), H$_3$PO$_4$ (85\%) and H$_2$O$_2$ (30\%) in the volume ratio $220 : 110 : 3 : 5$. We add the chemicals in that order to the solution and let it stand for around 10\,min at room temperature (RT) under agitation with a magnetic stirrer. We then etch at RT and remove around 100 to 200\,nm of material while continuing to agitate, checking the etch depth with a surface profilometer. The etching process is suspended by rinsing in DI water for 20\,s. The etch depth is chosen such that we are sure to have etched past the second, i.e., the lower, of the two Al(Ga)Sb confinement layers.
		\item Strip the photoresist in warm acetone and IPA ($\SI{50}{\celsius}$).  
	\end{itemize}
	\item Deposition of 25\,nm of Al$_2$O$_3$ at $\SI{150}{\celsius}$ using atomic layer deposition \\
	\item Definition of local gates with lift-off process:
	\begin{itemize}
		\item Perform dehydration bake for 120\,s at $\SI{120}{\celsius}$.
		\item Spin PMMA 50 K 4\% 1 : 1 in anisole at 1000\,rpm for 1\,s by accelerating for 1\,s from standstill, then accelerating for 5\,s to 5000\,rpm and holding for 45\,s. 
		\item Perform prebake for 240\,s at $\SI{180}{\celsius}$.
		\item Spin PMMA 950 K 6\% in anisole using the same settings as before.
		\item Perform prebake for 240\,s at $\SI{180}{\celsius}$. The total resist thickness is around 400\,nm.
		\item Perform electron-beam lithography (30\,keV beam energy, Raith 150 system), then develop for 30\,s in MIBK/IPA in 1 : 3 ratio at RT and rinse in IPA for 30\,s.
		\item Evaporate Ti/Au with thicknesses 10 and 130\,nm, respectively, using electron-beam evaporation, where the final 10\,nm of Au are evaporated at a tilt angle of $\SI{20}{\degree}$.
		\item Perform lift-off in DMSO at $\SI{80}{\celsius}$. 
	\end{itemize}
	\item Deposition of 25\,nm of Al$_2$O$_3$ at $\SI{150}{\celsius}$ using atomic layer deposition \\
	\item Definition of top gate with lift-off process
	\begin{itemize}
		\item Perform dehydration bake for 120\,s at $\SI{120}{\celsius}$
		\item Spin photoresist AZ 5214E at 3000\,rpm for 3\,s by accelerating for 3\,s from standstill, then accelerating for 5\,s to 5000\,rpm and holding for 60\,s. 
		\item Perform prebake for 110\,s at $\SI{60}{\celsius}$. The resulting photoresist thickness is around $\SI{1}{\micro\meter}$. 
		\item Perform exposure of the top gate pattern using contact printing with a quartz mask in a Karl S\"{i}ss MJB3 mask aligner (400\,nm 350 W Hg light source). The exposure lasts 3.5\,s.
		\item Perform reversal bake for 120\,s at $\SI{120}{\celsius}$.
		\item Perform flood exposure for 20\,s in the mask aligner, then develop in Microposit MF319 for 30\,s and rinse in DI water for 20\,s.
		\item Perform optional postbake for 60\,s at $\SI{120}{\celsius}$.
		\item Evaporate Ti/Au with thicknesses 10 and 130\,nm, respectively, using electron-beam evaporation, where the final 10\,nm of Au are evaporated at a tilt angle of $\SI{20}{\celsius}$.
		\item Perform lift-off in DMSO at $\SI{80}{\celsius}$. 
	\end{itemize}
	\item Definition of ohmic contacts with lift-off process:
	\begin{itemize}
		\item Perform dehydration bake for 120\,s at $\SI{120}{\celsius}$.
		\item Spin photoresist AZ 5214E at 3000\,rpm for 3\,s by accelerating for 3\,s from standstill, then accelerating for 5\,s to 5000\,rpm and holding for 60\,s. 
		\item Perform prebake for 110\,s at $\SI{60}{\celsius}$. The resulting photoresist thickness is around $\SI{1}{\micro\meter}$. 
		\item Perform exposure of the ohmic contact pattern using contact printing with a quartz mask in a Karl S\"{u}ss MJB3 mask aligner (400\,nm 350 W Hg light source). The exposure lasts 3.5\,s.
		\item Perform reversal bake for 120\,s at $\SI{120}{\celsius}$.
		\item Perform flood exposure for 20\,s in the mask aligner, then develop in Microposit MF319 for 30\,s and rinse in DI water for 20\,s.
		\item Perform postbake for 60\,s at $\SI{120}{\celsius}$.
		\item Etch the oxide stack and the heterostructure down to the InAs QW using BOE, employing the InAs as the etch stop layer.
		\item Evaporate Ti/Au with thicknesses 50 and 200\,nm, respectively, using electron-beam evaporation. The contacts are not annealed to prevent shorts to the back gate.
		\item Perform lift-off in DMSO at $\SI{50}{\celsius}$. 
	\end{itemize}
	\item Gluing of the sample into a chip carrier using PMMA or silver epoxy and bonding.
\end{enumerate}

\subsection{Device details}
We have investigated a total of three working devices, as summarized in Table 1 below. All devices have a width $W = \SI{5}{\micro\meter}$, and the distance between the junction voltage probes is $\SI{7}{\micro\meter}$ (Extended Data Fig.\ 2). Devices A and B are from the same wafer, and therefore share the same heterostructure. Device C is from a different wafer, and its heterostructure is identical apart from a decrease in the thickness $d_\mathrm{InAs}$ of the InAs QW by 1 nm. The consequence of decreasing $d_\mathrm{InAs}$is a reduction of the band inversion, and therefore of the degree of hybridization. All three devices show oscillations in the $\textit{pn'p}$ configuration, which we attribute to Fabry–-P\'{e}rot interference. Device A is presented in the main text, and B and C in the Supplementary Information.
\begin{table}[h]
	\centering
	\begin{tabular}{ | c | c | l|}
		\hline
		Device ID & QW thicknesses (nm) & Lithographic $L$ (nm) \\ \hline
		A & $d_\mathrm{GaSb}/d_\mathrm{InAs} = 8/13.5$ & $L = 320, 400, 540$ \\ \hline
		B & $d_\mathrm{GaSb}/d_\mathrm{InAs} = 8/13.5$ & $L = 530, 700$\\ \hline
		C & $d_\mathrm{GaSb}/d_\mathrm{InAs} = 8/12.5$ & $L = 320, 400, 540$\\
		\hline
	\end{tabular}
	\caption{Table of investigated devices.}
\end{table}

\subsection{Measurement setup}
All electronic transport experiments are conducted in a pumped 4He cryostat (also known as a variable temperature insert) at a base temperature of 1.3 K, or in a 3He/4He dilution refrigerator with a base temperature of 50 mK. The dc current $I = 50\text{–-}100$\,nA injected into the device (see Fig.~1a) is produced by applying a constant dc voltage over a series resistor which is orders of magnitude larger than the total resistance of the device and the cabling in the cryostat. A dc voltage $V_\mathrm{tg}$ is applied to the top gate, and a dc voltage $V_\mathrm{lg}$ together with an ac modulation voltage $\delta V_\mathrm{lg}$ is applied to the local gate under investigation using a lock-in. The frequency of this ac modulation is 31–-35 Hz and the amplitude is typically 10–-15 mV RMS.  The unused local gates and the back gate are grounded. The junction voltage $V_\mathrm{j} + \delta V_\mathrm{j}$ arising due to the current flow is given by  
\begin{equation}
V_\mathrm{j} = R_\mathrm{j}I \quad \mathrm{and} \quad \delta V_\mathrm{j} = \dfrac{\mathrm{d}R_\mathrm{j}}{\mathrm{d}V_\mathrm{lg}} \delta V_\mathrm{lg} I,
\end{equation}
where $R_\mathrm{j}$ is the dc junction resistance and $\mathrm{d}R_\mathrm{j}/\mathrm{d}V_\mathrm{lg}$ is the differential junction resistance which is sensitive to changes in the transmission of the region (cavity) below the local gate only. The dc component $V_\mathrm{j}$ is recorded with a multimeter, whereas the ac component $\delta V_\mathrm{j}$ is recorded with a lock-in. The four-terminal differential resistance is obtained by dividing $\delta V_\mathrm{j}$  with the ac modulation voltage and the input current, $\mathrm{d}R_\mathrm{j}/\mathrm{d}V_\mathrm{lg} = \delta V_\mathrm{j}/\delta V_\mathrm{lg} I$. The oscillations in the resistance, which are the focus of this work, are visible in both dc and ac components. However, we prefer to show the ac signal, as the oscillations are clearer there and the slowly varying background due to the changing resistance of the leads is automatically suppressed.

\section{Theoretical analysis}
\subsection{Model parameters}
To model the experiment, we use a minimal BHZ model, see equation (1). The parameters used throughout the presented theoretical analysis are the following: $\mathcal{A} = 32$\,meV nm, $\mathcal{B} = 400$\,meV nm$^2$, $\mathcal{M}_0 = -30$\,meV, $\mathcal{M}_2 = 700$\,meV nm$^2$, $L = 320$\,nm, $W = \SI{5}{\micro\meter}$.

\subsection{Theoretical modeling of Fabry–P\'{e}rot interference}
We employ a scattering matrix approach to calculate the transmission through the device shown in Fig. 1a from the main text. We approximate the device to be infinite in $y$-direction, i.e., we take the momentum $k_y$ to be a good quantum number. Due to the different chemical potentials in the cavity and the leads, effective barriers emerge at the cavity-lead interfaces and the system entails scattering of waves at these barriers. Since the scattering at the interfaces preserves the momentum $k_y$, we treat the two-dimensional scattering problem as a set of independent one- dimensional channels with different momenta $k_y$. As a result of the Sombrero hat band structure (see Figs. 2a, b and c), for each energy located within the region of band overlap there exists a region of momenta $k_y$ where two incident waves are present, one of them being electron-like and the other hole-like (see Extended Data Fig.\ 3).

We work in the limit of infinitely sharp p-n interfaces, although preliminary studies show that the oscillations are also present if the $p$-$n$ interfaces are broadened. In Extended Data Fig.\ 3, the scattering problem for one $k_y$-channel is illustrated. At every interface, the scattering matrix is equal to
\begin{equation}
\hat{S}_{L/R} = 
\begin{bmatrix}
r_{L/R} & t'_{L/R}\\
t_{L/R} & r'_{L/R}
\end{bmatrix},
\end{equation}
where the subscript L/R denotes the left/right interface and $r, r', t, t'$  are $2\times2$ matrices describing reflection and transmission amplitudes for electron-like and hole-like waves impinging on the interface from left ($r, t$) and right ($r’,  t’$) with the following structure:
\begin{equation}
\begin{aligned}
r = 
\begin{bmatrix}
r_{ee} & r_{eh} \\
r_{he} & r_{hh}
\end{bmatrix},
\quad
t = 
\begin{bmatrix}
t_{ee} & t_{eh} \\
t_{he} & t_{hh}
\end{bmatrix},
\\
r' = 
\begin{bmatrix}
r'_{ee} & r'_{eh} \\
r'_{he} & r'_{hh}
\end{bmatrix},
\quad
t' = 
\begin{bmatrix}
t'_{ee} & t'_{eh} \\
t'_{he} & t'_{hh}
\end{bmatrix}.
\end{aligned}
\end{equation}

We can write down the scattering equations for the full $pn'p$ junction, 
\begin{equation}
\begin{aligned}
\begin{pmatrix}  \vec{\Psi}^L_{-} \\ \vec{\Psi}^C_{+}  \end{pmatrix} \bigg\lvert_{x=0} &= \hat{S}_L \begin{pmatrix}  \vec{\Psi}^L_{+} \\ \vec{\Psi}^C_{-}  \end{pmatrix} \bigg\lvert_{x=0},\\
\begin{pmatrix}  \vec{\Psi}^C_{-}(x=0) \\ \vec{\Psi}^C_{+}(x=L)  \end{pmatrix} &= \hat{T} \begin{pmatrix}  \vec{\Psi}^C_{-}(x=L) \\ \vec{\Psi}^C_{+}(x=0)  \end{pmatrix},\\
\begin{pmatrix}  \vec{\Psi}^C_{-} \\ \vec{\Psi}^R_{+}  \end{pmatrix} \bigg\lvert_{x=L} &= \hat{S}_R \begin{pmatrix}  \vec{\Psi}^C_{+} \\ \vec{\Psi}^R_{-}  \end{pmatrix} \bigg\lvert_{x=L},		
\end{aligned}
\end{equation}
where C denotes the cavity and the subscripts $+/-$ denote waves moving to the right/left, i.e., waves with positive/negative velocities. In general, $\vec{\Psi}_{\pm}^{j} = [\, \psi^{j}(\pm|k_e|) \quad \psi^{j}(\mp|k_h|) \,]^T$ is a two- component vector with spinor components that denote waves with electron-like dispersion (marked with red circles in Extended Data Fig.\ 3) and waves with hole-like dispersion (blue circles in Extended Data Fig.\ 3). The matrix $\hat{T}$ assigns the dynamical phase for the waves due to their free propagation, and is equal to
\begin{equation}
\begin{aligned}
\hat{T} = 
\begin{bmatrix}
e^{i |k_e| L} & 0 & 0 & 0\\
0 & e^{-i |k_h| L} & 0 & 0\\
0 & 0 & e^{i |k_e| L} & 0\\
0 & 0 & 0 & e^{-i |k_h| L}\\
\end{bmatrix}
\equiv
\begin{bmatrix}
\hat{\phi} & 0 \\ 
0 & \hat{\phi}
\end{bmatrix}.
\end{aligned}
\end{equation}
Now that we have set up the scattering problem in terms of scattering processes at each junction individually, we can proceed in determining the total transmission through the device. Let us write down the equations that connect the incident states from the left lead to the transmitted waves in the right lead,
\begin{equation}
\begin{aligned}
\begin{pmatrix}
\vec{\Psi}_{-}^{L}\\
\vec{\Psi}_{+}^{R}
\end{pmatrix}
&=
\hat{S}
\begin{pmatrix}
\vec{\Psi}_{+}^{L}\\
\vec{\Psi}_{-}^{R}
\end{pmatrix}
\end{aligned}
\end{equation}
where $\hat{s}$ is the scattering matrix of the whole $pn’p$ junction,
\begin{equation} 	
\begin{aligned}
\hat{S} &= 
\begin{bmatrix}
\tilde{r} & \tilde{t}'\\
\tilde{t} & \tilde{r}'
\end{bmatrix}.
\end{aligned}
\end{equation}
In the experiment, current is injected only from the left lead and so we set $\vec{\Psi}_{-}^{R} = 0$. The final solution for the waves transmitted in the right lead, $\vec{\Psi}_{+}^{R}$, is
\begin{equation} 	
\begin{aligned}
\vec{\Psi}_{+}^{R} = \tilde{t} \, \vec{\Psi}_{+}^{L}.
\end{aligned}
\end{equation}
The matrix $\tilde{t}$ can be easily obtained from the matrices $\hat{S}_L$ , $\hat{S}_R$ and $\hat{\phi}$. After considering time-
reversal symmetry, $\hat{S}_{L/R}^T=\hat{S}_{L/R}$, and inversion symmetry around the middle of the cavity ($x = L/2$), $\hat{S}_R(E, k_y)=\sigma_x \hat{S}_L(E, -k_y) \sigma_x$, where $\sigma_x$ is a Pauli matrix, we obtain
\begin{equation} 	
\begin{aligned}
\tilde{t} = \overline{t}^T \hat{\phi} \, \left( \hat{\mathbf{1}}_{2\times2} - r\, \hat{\phi}\, \overline{r}\, \hat{\phi} \right)^{-1} \, t \, ,
\end{aligned}
\end{equation}
with $\{ \overline{r}(k_y), \overline{t}(k_y) \} = \{ r(-k_y), t(-k_y) \}$.
Finally, the total transmission through our Fabry–-P\'{e}rot interferometer, as shown in Fig.\ 2e, is
\begin{equation}
T(E', k_y) = \mathrm{tr} \, \tilde{t}^{\dagger} \tilde{t} \, .
\end{equation}

The solutions of the Schr\"{o}dinger equation in both cavity and leads are plane waves with $\vec{k}(E_F)$
momenta, which are obtained from the eigenvalues of the Hamiltonian in equation (1) from the main text, and depend on the Fermi energy inside the cavity and the leads. For a one-dimensional problem, the projection of the momentum wavevector in $x$-direction depends both on the Fermi energy and the incident momentum in y-direction, i.e., $k_x^{e/h}(E_F, k_y)$, where $e/h$ denotes electron-like/hole-like states (see Extended Data Fig.\ 3).

We concentrate on one interface—a $n'p$ junction, i.e., the system consists of an $n$-type cavity and a $p$-type lead. The cavity and the lead are semi-infinite in $–x$ and $+x$ directions, respectively. Since the coefficients that enter the overall scattering matrix of a Fabry–-P\'{e}rot interferometer depend on the reflections inside the cavity, we consider the incident wave from the cavity to the lead. The wave function in the cavity is then given by the sum of incident and outgoing waves
\begin{equation} 	
\begin{aligned}
\vec{\Psi}^{cavity}(E', k_y, x)= 
\begin{pmatrix}
e^{i|k^e_x| x} \psi(|k^e_x|, k_y) +a_{ee} \, e^{-i|k^e_x| x} \psi(-|k^e_x|, k_y) + a_{he} \, e^{i|k^h_x| x} \psi(|k^h_x|, k_y)  \\
e^{-i|k^h_x| x} \psi(-|k^h_x|, k_y) + a_{hh} \, e^{i|k^h_x| x} \psi(|k^h_x|, k_y)- a_{eh} \, e^{-i|k^e_x| x} \psi(-|k^e_x|, k_y)
\end{pmatrix},
\end{aligned}
\end{equation}
while inside the lead, there are only outgoing transmitted waves:
\begin{equation} 	
\begin{aligned}
\vec{\Psi}^{lead}(E, k_y, x)= 
\begin{pmatrix}
b_{ee} e^{i|k^e_x| x} \psi(|k^e_x|, k_y) +  b_{he} \, e^{-i|k^h_x| x}\psi(-|k^h_x|, k_y) \\
b_{eh} e^{i|k^e_x| x} \psi(|k^e_x|, k_y) +  b_{hh} \, e^{-i|k^h_x| x}\psi(-|k^h_x|, k_y) 
\end{pmatrix},
\end{aligned}
\end{equation}
where $ \psi(k_x, k_y)$ is a momentum-dependent spinor, and $k_x^{e/h}$ is the momentum in $x$-direction which depends on the Fermi energy $E$ and $k_y$. At the junction between the cavity and the lead, the wave function and its derivative must be continuous,
\begin{equation} 	
\begin{aligned}
\vec{\Psi}^{cavity}(E', k_y)|_{x=0} &= \vec{\Psi}^{lead}(E, k_y) |_{x=0}\\
\partial_x \, \vec{\Psi}^{cavity}(E', k_y)|_{x=0} &= \partial_x \, \vec{\Psi}^{lead}(E, k_y)|_{x=0},
\end{aligned}
\end{equation}
which then yields eight equations, allowing us to calculate all the coefficients $a_{ij}$ and $b_{ij}$. To obtain the final reflection and transmission coefficients that enter the scattering matrices at the left and right interfaces, i.e., the $pn'$ and $n'p$ junctions, we need to multiply $a_{ij}$ and $b_{ij}$ with the respective velocities in $x$-direction
\begin{equation} 	
\begin{aligned}
r_{ij} = a_{ij} \sqrt{ \frac{v_i}{v_j} }, \qquad t_{ij} = b_{ij} \sqrt{ \frac{v_i}{v_j} },
\end{aligned}
\end{equation}
where $v_i=\mathrm{d} E / \mathrm{d} k_x$.

To obtain the interference condition in Eq. (2) of the main text, it is sufficient to look at the factor $\left( \hat{\mathbf{1}}_{2\times2} - r\, \hat{\phi}\, \overline{r}\, \hat{\phi} \right)^{-1} $. Assuming solely $eh$ and $he$ scattering processes ($r_{ee} = r_{hh} = 0$), this factor becomes
\begin{equation} 	
\begin{aligned}
\left( \hat{\mathbf{1}}_{2\times2} - r\, \hat{\phi}\, \overline{r}\, \hat{\phi} \right)^{-1}&= 
\left( 
\hat{\mathbf{1}} -
\begin{bmatrix}
0 & r_{eh}\\
r_{he} & 0
\end{bmatrix}
\begin{bmatrix}
e^{i |k_e| L} & 0\\
0 & e^{-i |k_h| L}
\end{bmatrix}
\begin{bmatrix}
0 & \overline{r}_{eh}\\
\overline{r}_{he} & 0
\end{bmatrix}
\begin{bmatrix}
e^{i |k_e| L} & 0\\
0 & e^{-i |k_h| L}
\end{bmatrix}
\right)^{-1}\\
&=
\left( 
\hat{\mathbf{1}} -
\begin{bmatrix}
r_{eh}\overline{r}_{he} e^{i \Delta k L} & 0\\
0 & r_{he}\overline{r}_{eh} e^{i \Delta k L}
\end{bmatrix}
\right)^{-1}\\
&=
\begin{bmatrix}
\frac{1}{1-r_{eh}\overline{r}_{he} e^{i \Delta k L}} & 0\\
0 & \frac{1}{1-r_{he}\overline{r}_{eh} e^{i \Delta k L}}
\end{bmatrix},
\end{aligned}
\end{equation}
where $\Delta k \equiv \Delta k_{eh(he)}$. The coefficients $r_{he}$ and $r_{eh}$ are in general complex numbers, and may carry an additional phase. Therefore, we rewrite the above expression as
\begin{equation} 	
\begin{aligned}
\begin{bmatrix}
\frac{1}{1-|r_{eh}||\overline{r}_{he}| e^{i \varphi}} & 0\\
0 & \frac{1}{1-|r_{he}||\overline{r}_{eh}| e^{i \varphi}}
\end{bmatrix} \, .	
\end{aligned}
\end{equation}
The total phase of the oscillations is then $\varphi(E, k_y) = \Delta k L + \delta$, where the phase $\delta = \delta_{eh} + \delta_{he}$ comes from the reflection coefficients. In Extended Data Fig.\ 4, we plot $\delta_{eh}$ and $\delta_{he}$ as a function of the Fermi energy $E$ inside the cavity and incident $k_y$ momenta. Since the phases only weakly depend on $E$ and $k_y$, i.e., $|\delta(E, k_y)| \approx 2\pi$, we can neglect them and write the condition for constructive interference as $\varphi(E_m, k_y) = \Delta k L + \delta \approx \Delta k L = 2m\pi$.

The magnetic field dependence was calculated using the open source Python package Kwant~\cite{groth_kwant:_2014}. We have used a Savitzky–-Golay filter on the obtained numerical data with a window of $0.5$\,meV to mimic the effect of smoothing that temperature has in the experiment.

\setcounter{figure}{0}    
\renewcommand{\figurename}{Extended Data FIG.}
\makeatletter
\@fpsep\textheight
\makeatother

\begin{figure*}[p!]
	\includegraphics[]{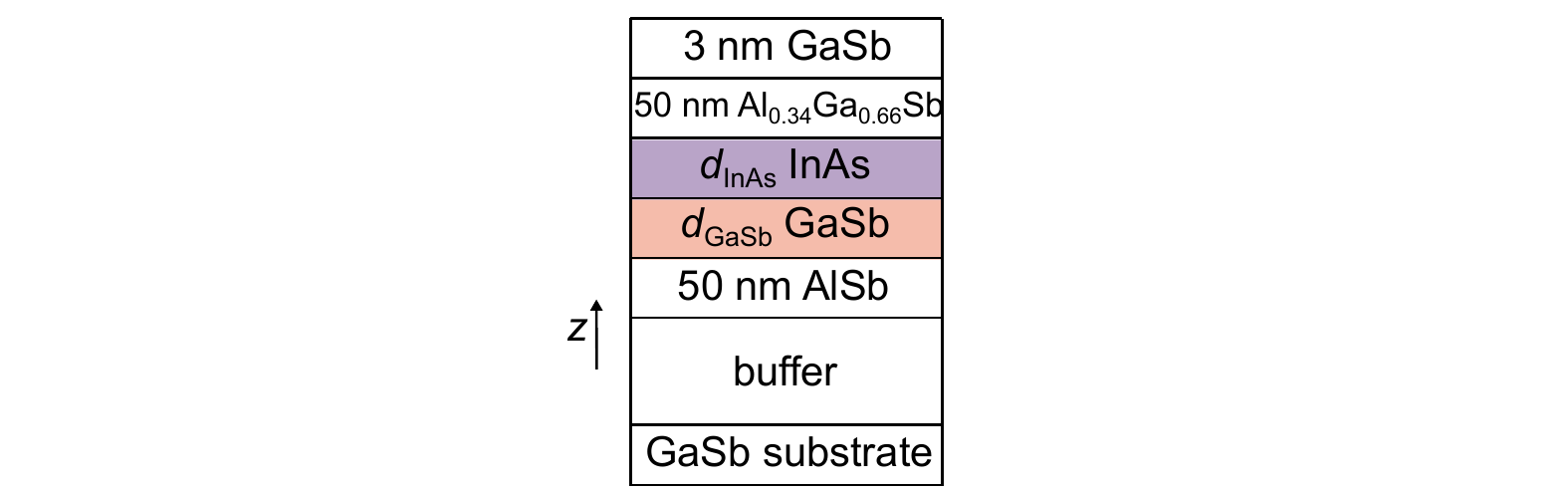}
	\caption{{\bf Heterostructure composition.} Layer sequence of the InAs/GaSb heterostructures used in this work. The growth direction is along the $z$-axis. The thicknesses of the InAs and GaSb quantum wells are $d_\mathrm{InAs}$ and $d_\mathrm{GaSb}$, respectively.} 
	\label{figExt1}
\end{figure*}

\begin{figure*}[p!]
	\includegraphics[]{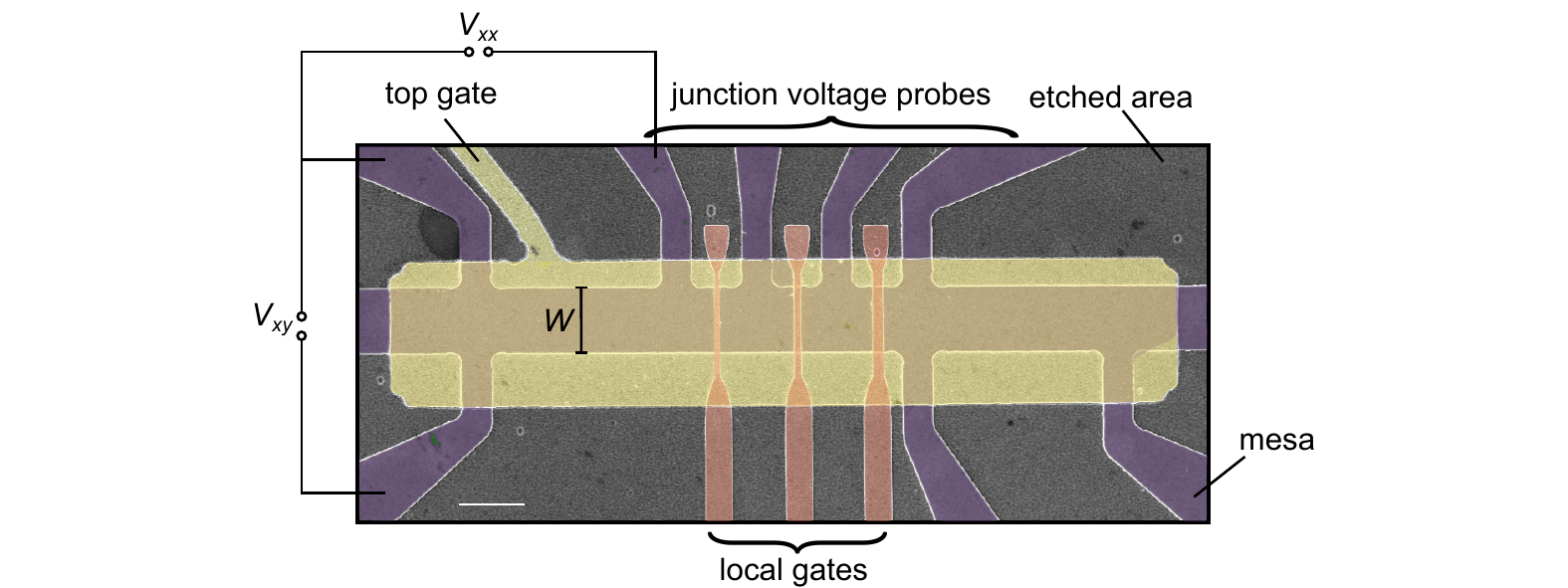}
	\caption{{\bf False-color image of a completed device.} Scanning electron micrograph of a typical device with a total of three local gates. The scale bar represents $\SI{5}{\micro\meter}$. The voltages $Vxx$ and $Vxy$ are used to determine the resistivities $\rho_{xx}$ and $\rho_{xy}$ in the bulk region covered by the top gate exclusively.} 
	\label{figExt2}
\end{figure*}

\begin{figure*}[p!]
	\includegraphics[]{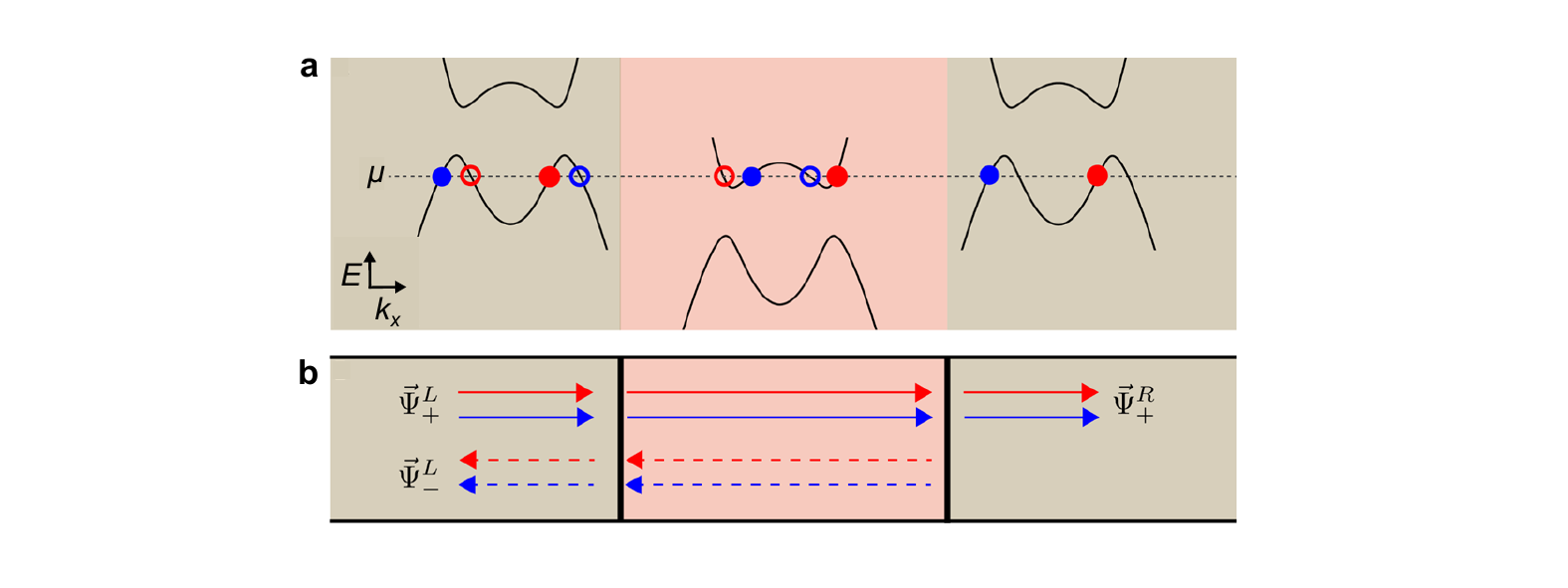}
	\caption{{\bf Fabry–-P\'{e}rot interference in a one-dimensional wire.} \textbf{a}, Band structure in the three regions of a $pn’p$ junction. The leads surround the cavity, located in the center. The dashed horizontal line represents the electrochemical potential $\mu$. Full blue (red) circles represent hole-like (electron-like) states moving to the right, while empty blue (red) circles represent hole-like (electron-like) states moving to the left. \textbf{b}, Sketch of a Fabry–-P\'{e}rot interferometer for a one-dimensional channel. The colors are coded according to \textbf{a}, with red (blue) arrows denoting hole-like (electron-like) states moving to the right (solid lines) and left (dashed lines).} 
	\label{figExt3}
\end{figure*}

\begin{figure*}[p!]
	\includegraphics[]{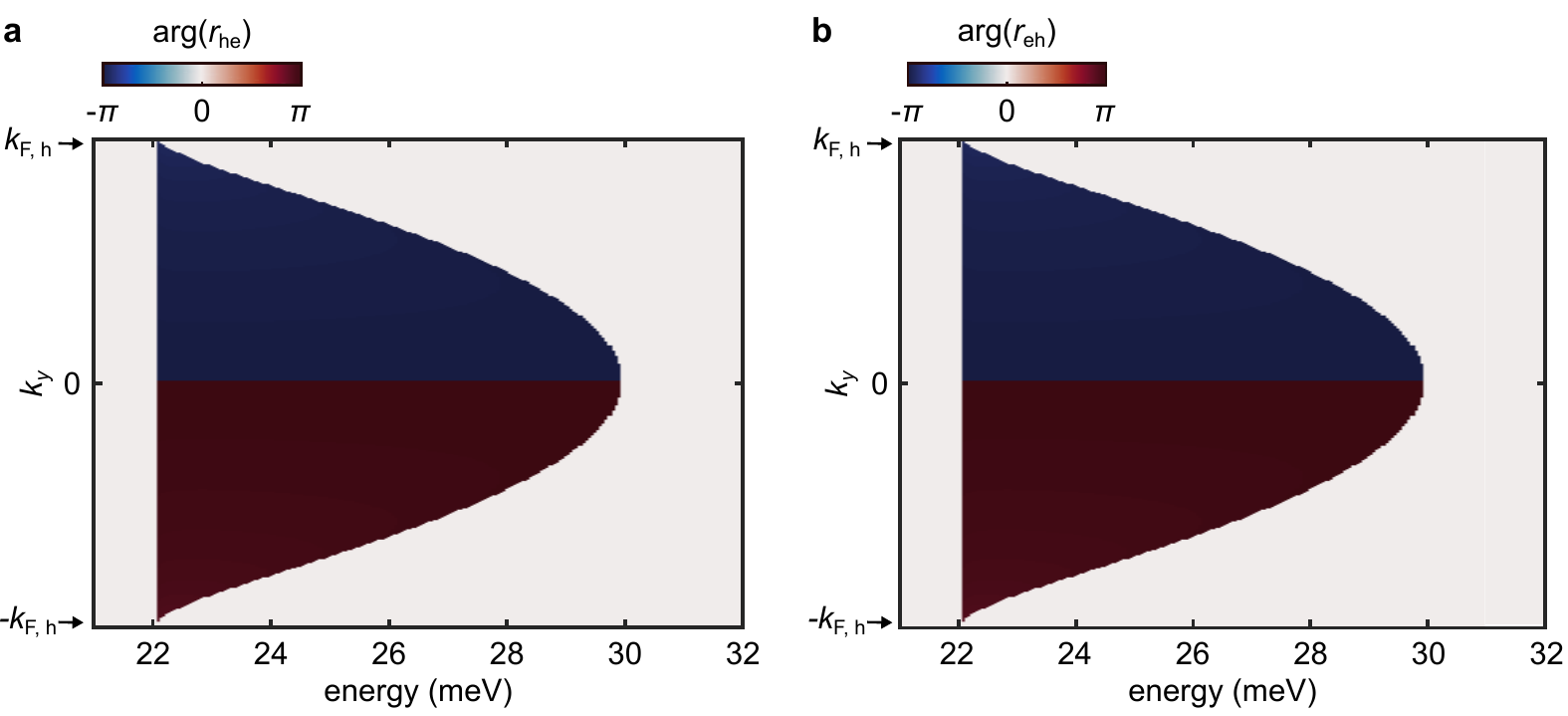}
	\caption{{\bf Phases of reflection coefficients.} \textbf{a}, The phase $\delta_\mathrm{he}$ of the reflection coefficient $r_\mathrm{he}$. \textbf{b}, The phase $\delta_\mathrm{eh}$ of the coefficient $r_\mathrm{eh}$.}
	\label{figExt4}
\end{figure*}

\setcounter{figure}{0}    
\renewcommand{\figurename}{Supplementary FIG. S}
\makeatletter
\@fpsep\textheight
\makeatother

\begin{figure*}[p!]
	\includegraphics[]{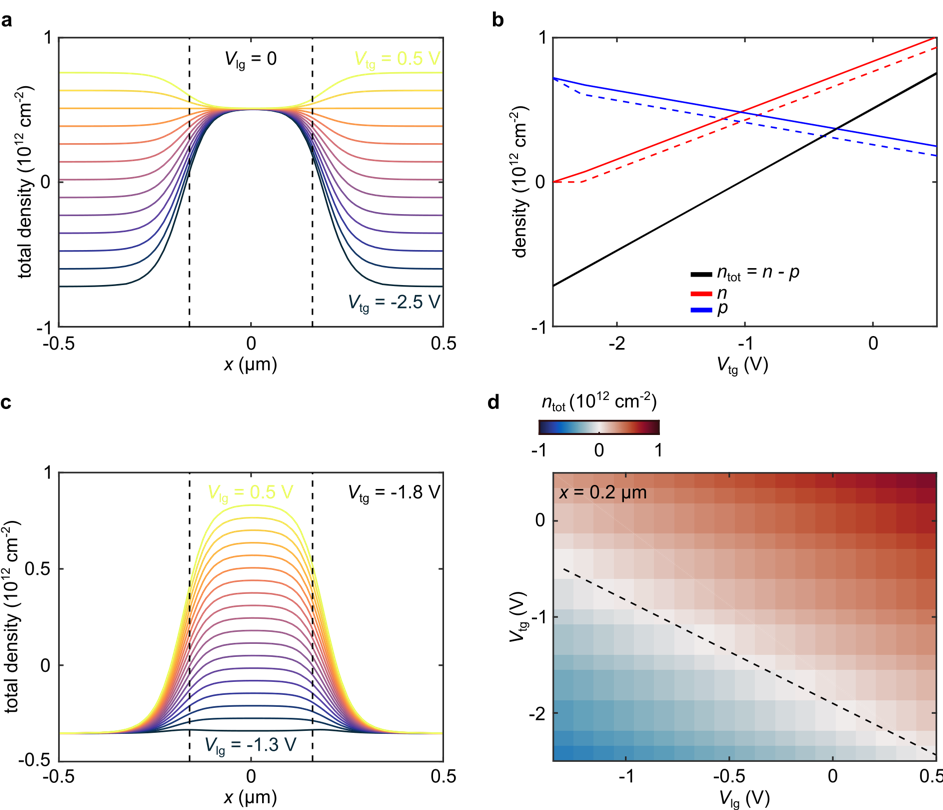}
	\caption{
		{\bf Electrostatic simulations.} 
		\textbf{a}, Simulated spatial dependence of the total density ntot below a local gate of length $L = 320$ nm for different values of $V_{\rm tg}$ at $V_{\rm lg} = 0$. The local gate is centered at $x = 0$, and the vertical dashed lines indicate its extent. 
		\textbf{b}, Simulated density (solid lines) as a function of $V_{\rm tg}$ below the top gate, i.e., in the outer device regions forming the leads to the locally gated regions. The total, electron and hole densities are denoted by $n_{tot}$, $n$ and $p$, respectively.  Also shown is the experimentally determined density (dashed lines), taken from Supplementary Fig. S2a. In the case of $n_{tot}$, there is an almost perfect agreement between simulation and experiment. 
		\textbf{c}, As in (\textbf{q}), but now for different values of $V_{\rm lg}$ at $V_{\rm tg} = -1.8$ V. 
		\textbf{d},, Simulated total density ntot as a function of $V_{\rm lg}$ and $V_{\rm tg}$ at $x = 0.2$ $\mu$m, at the transition between the density below the local gate of length $L = 320$ nm and the density in the outer region, which is influenced by the top gate exclusively. The dashed line follows $n_{tot} = 0$.  
	}
\end{figure*}

\begin{figure*}[p!]
	\includegraphics[]{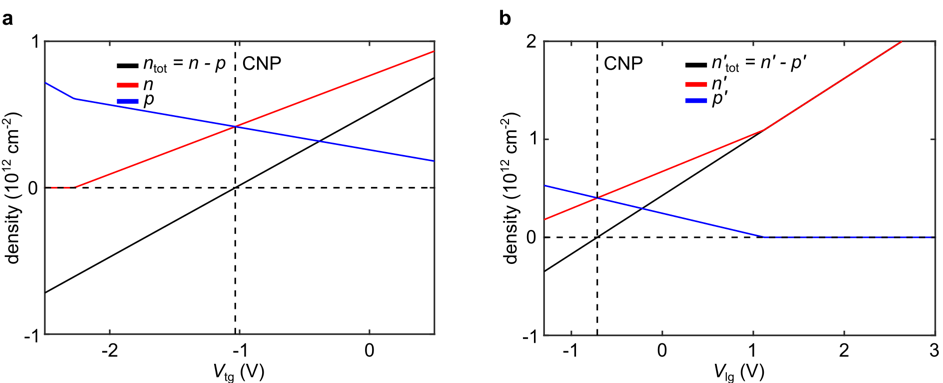}
	\caption{
		{\bf Density dependence below the top gate and in the cavity.} 
		\textbf{a}, Density as a function of $V_{\rm tg}$ below the top gate, i.e., in the outer device regions forming the leads to the cavity. The total, electron and hole densities are denoted by $n_{tot}$, $n$ and $p$, respectively.
		\textbf{b}, Density as a function of $V_{\rm lg}$ in the cavity below the local gate of length $L = 320$ nm. The total, electron and hole densities are denoted by $n'_{tot}$, $n'$ and $p'$, respectively.
	}
\end{figure*}

\begin{figure*}[p!]
	\includegraphics[]{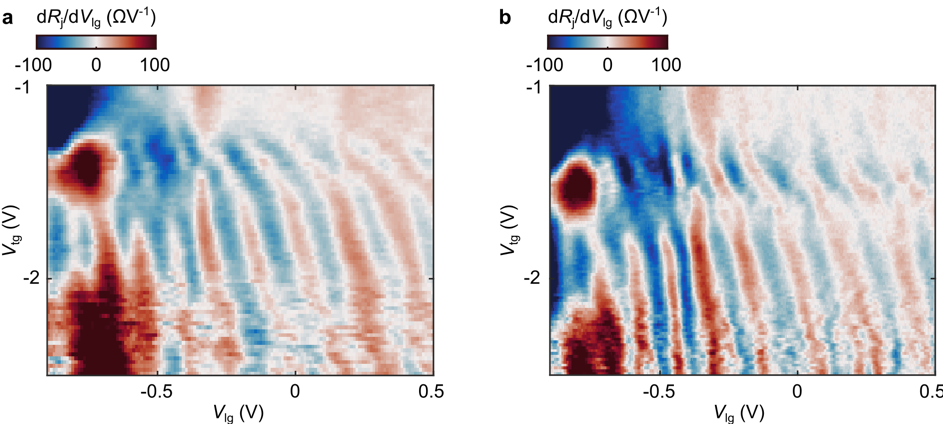}
	\caption{
		{\bf Reproducibility of the oscillations in different cooldowns.} 
		\textbf{a}, Differential junction resistance $d R_j/d V_{\rm lg}$ for the cavity with local gate length $L = 320$ nm as a function of $V_{\rm lg}$ and $V_{\rm tg}$ at 1.3 K in a first cooldown, presented in a reduced range in order to appreciate the interference phenomenon more clearly. 
		\textbf{b}, As in \textbf{a}, but in a second cooldown, performed several months after the first.  The device was stored in vacuum at RT in between the cooldowns.
	}
\end{figure*}

\begin{figure*}[p!]
	\includegraphics[]{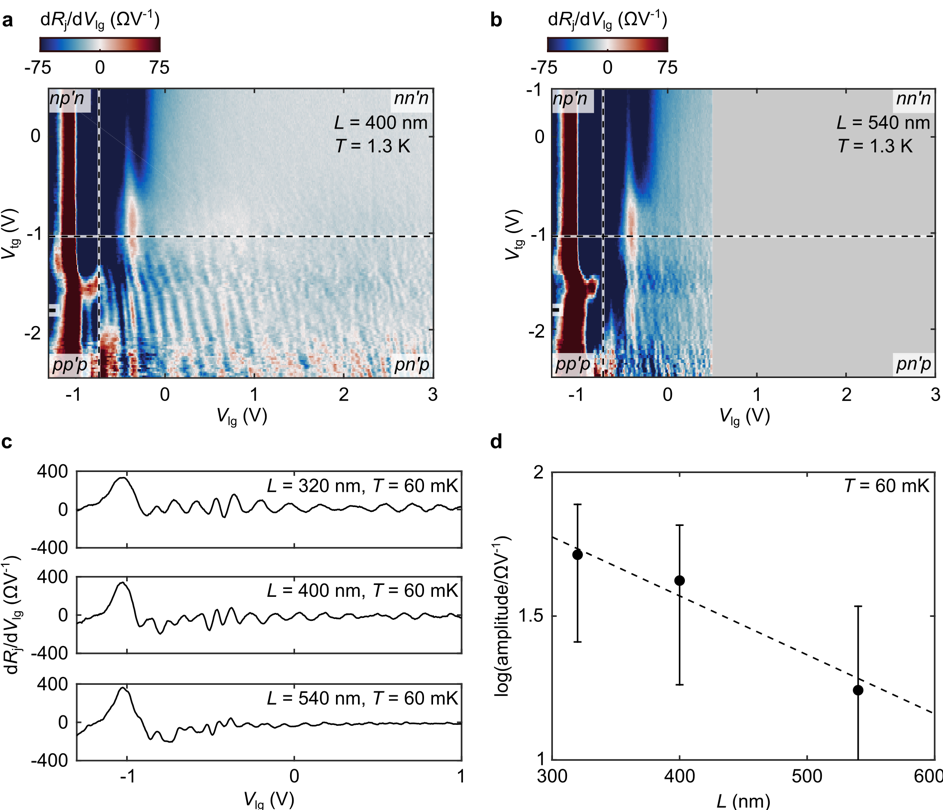}
	\caption{
		{\bf Local gate length dependence. }
		\textbf{a}, Color map of $d R_j/d V_{\rm lg}$ as a function of $V_{\rm lg}$ and $V_{\rm tg}$ at 1.3 K for the cavity with local gate of nominal length $L = 400$ nm. The resulting phase diagram is subdivided into quadrants according to the charge carrier configuration. The horizontal and vertical dashed lines mark the CNPs of the leads and cavity, respectively. 
		\textbf{b}, As in \textbf{a}, but for the local gate of nominal length $L = 540$ nm. Note the reduced range in $V_{\rm lg}$; no data points were recorded in the shaded area. 
		\textbf{c}, Differential junction resistance $d R_j/d V_{\rm lg}$ for all three cavities of device A at 60 mK as a function of the respective gate voltage $V_{\rm lg}$ at $V_{\rm tg} = -1.8$ V. 
		\textbf{d}, Semi-logarithmic plot of the average oscillation amplitude versus length L for device A, determined from c in the interval $-0.85 V < V_{\rm lg} < 1$ V. The error bars indicate a symmetric interval with width equal to two standard deviations.  The dashed line is an inverse-variance weighted best fit of the form $A_0 \exp(-2L/l_{\varphi})$, where $l_{\varphi}$ is the phase coherence length.    
	}
\end{figure*}

\begin{figure*}[p!]
	\includegraphics[]{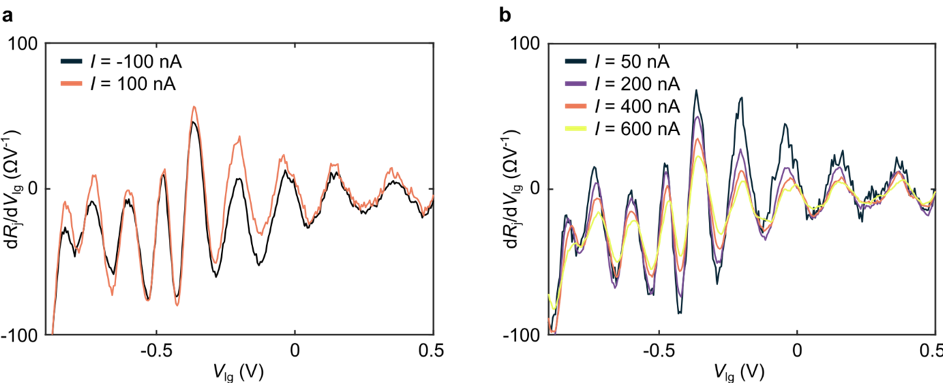}
	\caption{
		{\bf Bias current dependence. } 
		\textbf{a}, Differential junction resistance $d R_j/d V_{\rm lg}$ for the cavity with local gate length $L = 320$ nm at 1.3 K as a function of $V_{\rm lg}$ at $V_{\rm tg} = -1.8$ V. in the $pn'p$ configuration and in a reduced range in $V_{\rm lg}$, chosen so as to highlight the oscillations’ behavior under current reversal.
		\textbf{b}, Similar to \textbf{a}, but spanning a larger current range for a fixed bias direction.  
	}
\end{figure*}

\begin{figure*}[p!]
	\includegraphics[]{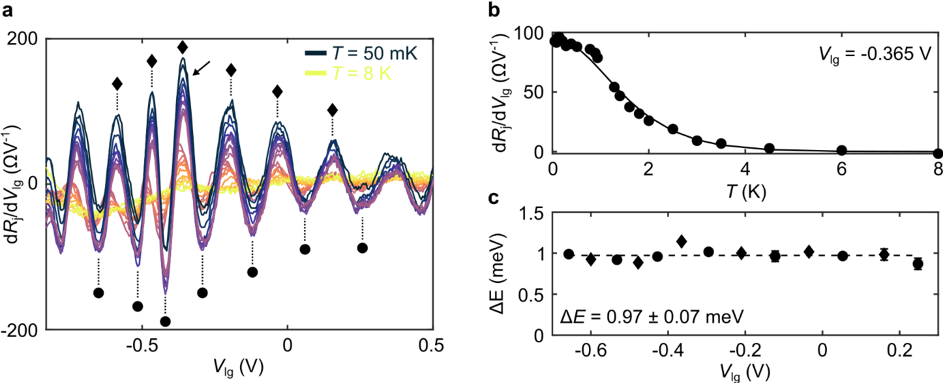}
	\caption{
		{\bf Temperature dependence } 
		\textbf{a}, Differential junction resistance $d R_j/d V_{\rm lg}$ for the cavity with local gate length $L = 320$ nm at 1.3 K as a function of $V_{\rm lg}$ at $V_{\rm tg} = -1.8$ V. in the $pn'p$ configuration and in a reduced range in $V_{\rm lg}$, chosen so as to highlight the oscillations’ behavior upon changing the temperature from 50 mK to 8 K. The dotted lines and symbols mark local minima (circles) and maxima (diamonds) used in the analysis presented in \textbf{c}.
		\textbf{b}, Fit of the temperature dependence of the local maximum at $V_{\rm lg} = -0.365$ V, indicated by the arrow in \textbf{a}, with a thermal damping function, as described in the text.
		\textbf{c}, Summary of the mode spacing $\Delta E$ in the cavity extracted from the local minima and maxima marked in \textbf{a}. The mean and standard deviation obtained when factoring in all points is inset. The individual points have relative errors of around 5$\%$, such that the error bars are as large as or smaller than the point size. 
	}
\end{figure*}

\begin{figure*}[p!]
	\includegraphics[]{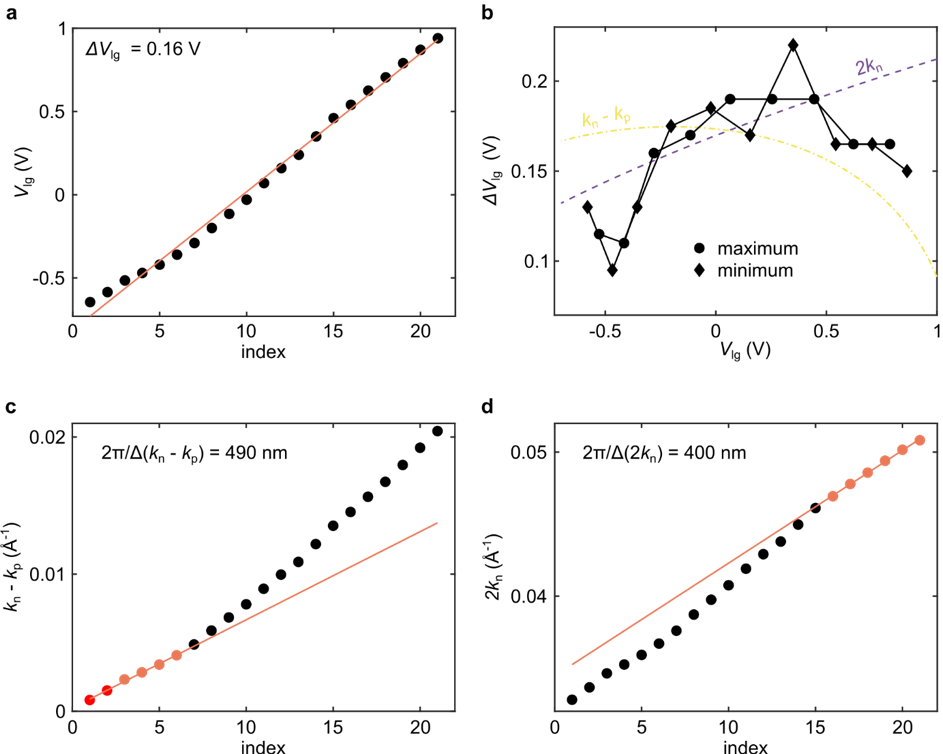}
	\caption{
		{\bf Oscillation period analysis. }
		\textbf{a}, Index plot showing the positions of minima and maxima in $d R_j/d V_{\rm lg}$ in the interval $-0.7 < V_{\rm lg} < 1$ V for the cavity with local gate length $L = 320$ nm, taken at constant   $V_{\rm tg} = -1.8$ V and 60 mK (trace in top panel of Supplementary Fig. S4c). The solid line is a best fit with a linear function through all data points, resulting in an average period $\Delta V_{\rm lg} = 0.16$ V.
		\textbf{b}, Evolution of the period $\Delta V_{\rm lg}$ as a function of $ V_{\rm lg}$, calculated by taking differences between adjacent maxima (circles) and minima (diamonds) in \textbf{a}. Also shown are expected $\Delta V_{\rm lg}$ trajectories if the oscillations were exactly periodic in $k_n - k_p$ (dash-dotted line), or in $2 k_n$ (dashed line). 
		\textbf{c}, Index plot of \textbf{a}, wherein the vertical axis is converted to $k_n - k_p$, together with a best fit through the first six points, resulting in an average period $\Delta (k_n - k_p)$.   
		\textbf{d}, As in \textbf{c}, but with the vertical axis converted to $2 k_n$, together with a best fit through the last six points, resulting in an average period $\Delta (2 k_n)$.
	}
\end{figure*}

\begin{figure*}[p!]
	\includegraphics[]{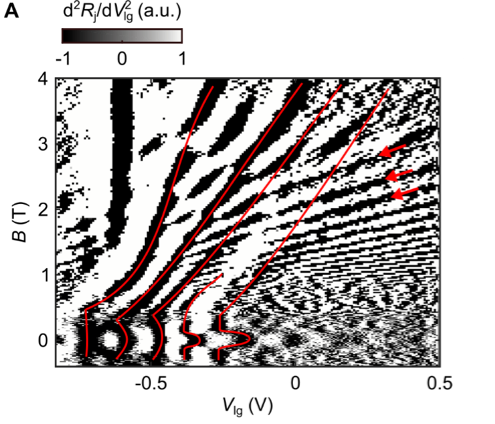}
	\caption{
		{\bf Extended magnetic field dependence. } 
		Color map of the numerical derivative of $d R_j/d V_{\rm lg}$  with respect to $V_{\rm lg}$ as a function of $V_{\rm lg}$ and $B$ for the cavity with local gate length $L = 320$ nm, taken at $V_{\rm tg} = -1.8$ V and 1.3 K. The arrows and solid lines are guides to eye.
	}
\end{figure*}

\begin{figure*}[p!]
	\includegraphics[]{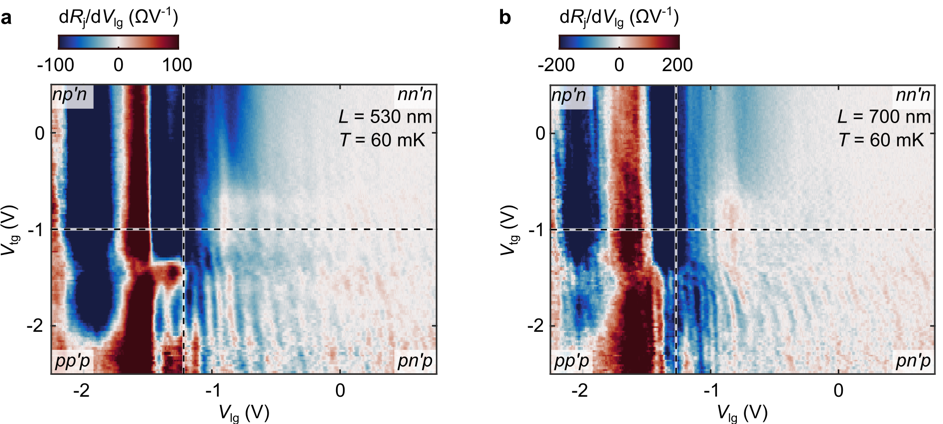}
	\caption{
		{\bf Quantum interference in device B. }
		\textbf{a}, Color map of $d R_j/d V_{\rm lg}$  in device B as a function of $V_{\rm lg}$ and $V_{\rm tg}$ at 60 mK for the cavity with nominal local gate length $L = 530$ nm. The resulting phase diagram is subdivided into quadrants according to the charge carrier configuration. The horizontal and vertical dashed lines mark the CNPs of the leads and cavity, respectively. 
		\textbf{b}, As in \textbf{a}, but for the local gate of nominal length $L = 700$ nm, also in device B.
	}
\end{figure*}

\begin{figure*}[p!]
	\includegraphics[]{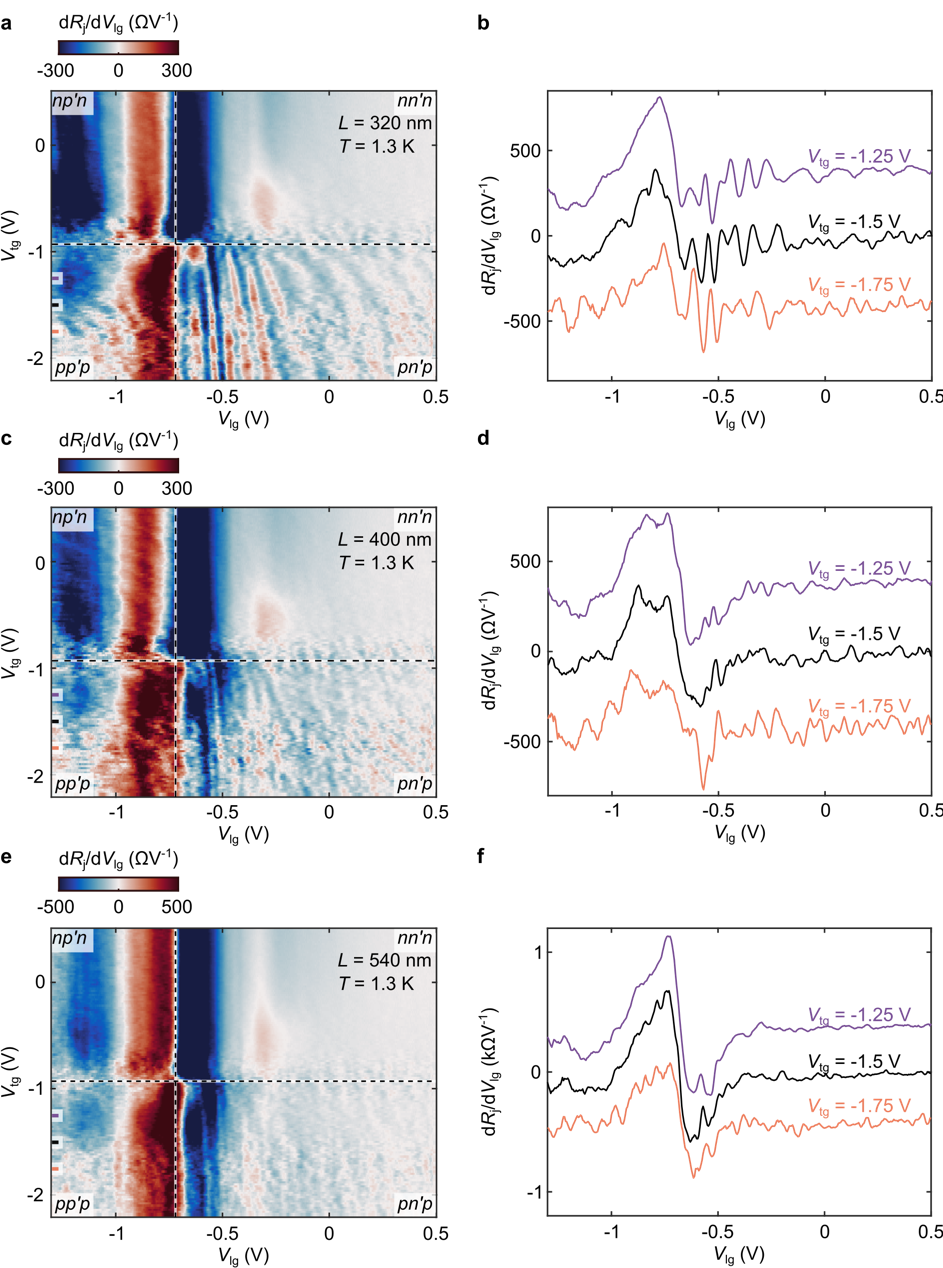}
	\caption{
		{\bf Quantum interference in device C.  }
		Resistance oscillations in device C, shown in color maps \textbf{a}, \textbf{c} and \textbf{e} as well as line cuts \textbf{b}, \textbf{d} and \textbf{f} which are offset vertically for clarity.
	}
\end{figure*}

\renewcommand{\thefigure}{Supplementary Figure S\arabic{figure}}
\bigskip
The points discussed in the following all pertain to device A, unless explicitly stated otherwise. 

\section{Electrostatic simulation of the gating}
We simulate the electrostatics of our experiment with the finite element method using COMSOL Multiphysics to better understand the combined operation of the local gates and the top gate. Our gate geometry comprises a single local gate of length $L = 320$ nm which is centered at $x = 0$, where $x$ is the direction of current flow, see. Fig.~1 of the main text and the coordinate system introduced therein, and a top gate. We faithfully incorporate all the relevant layers of the device with their respective thicknesses and dielectric constants into the model. The two quantum wells (QWs) are represented by two infinitely thin charged sheets, and we approximate the density of states within each QW by a constant value (effective mass approximation).

Supplementary Fig.~S1a depicts the effect of the top gate voltage $V_{\rm tg}$ on the total density $n_{\rm tot}$ below the local gate and in its immediate vicinity. The local gate voltage is fixed to zero, $V_{\rm lg} = 0$. The vertical dashed lines represent outlines of the local gate. We observe that the density at $x = 0$ is independent of $V_{\rm tg}$. However, due to fringe field effects the electric field generated by the top gate also influences the density below the local gate, and this influence increases away from $x = 0$. To characterize the sharpness of the transition we determine $x^*$, the position where $n_{\rm tot}$ is equal to the mean of the total density at $x = 0$ and the density in the bulk, i.e., sufficiently far away from the local gate. The quantity $x^*$ is therefore the half width at half maximum (HWHM) of $n_{\rm tot}(x)$ and it is a measure of the sharpness of the charge carrier confinement below the local gate, in the cavity. Here, we find that $x^* \approx 200$ nm for all $V_{\rm tg}$.

In order to verify that the simulation, as described above, accurately describes our device, we extract the bulk density in the region covered by the top gate only and compare it with experiment, as detailed in Supplementary Fig.~S1b. The solid lines are extracted from the simulation, and the dashed lines are taken from the experiment, from Supplementary Fig.~S2a, see also the next section for details. The agreement in $n_{\rm tot}$ is excellent (the curves fall on top of each other), and it is also satisfactory in the constituent electron and hole densities, $n$ and $p$, respectively. This overall consistency guarantees the validity of our model and creates confidence in drawing overarching conclusions from the simulation. Note that the simulation does not take the hybridization of bands into account, and this might explain the remaining discrepancies in $n$ and $p$.

Now we fix $V_{\rm tg}$ to a value at which we observe interference in the $pn'p$ configuration, and study the effect of $V_{\rm lg}$ on the density below the local gate. To this end, we set $V_{\rm tg} = -1.8$V and vary $V_{\rm lg}$, see Supplementary Fig.~S1c. The resulting simulated $n_{\rm tot}(x)$ profile is similar to what we saw before, i.e., as in Supplementary Fig.~S1a, but with the roles of the gates reversed. Again, we calculate the HWHM for all possible values of $V_{\rm lg}$ and find $x^* \approx 200$ nm everywhere.

While we conclude that the sharpness of the density profile is described by an approximately constant HWHM, $x^*$, for all combinations of $V_{\rm lg}$ and $V_{\rm tg}$, the density at a certain fixed position x changes with both gate voltages. Supplementary Fig.~S1d exemplifies this phenomenon. We recognize the negative cross-coupling between the local and top gates by the negative slope of the contour lines in the color map—decreasing $V_{\rm tg}$ in the $pn'p$ configuration, for instance, squeezes the cavity, and to compensate, i.e., to maintain the same $n'$, an increase in $V_{\rm lg}$ is necessary.

From the theory of quantum interference of partially reflected and transmitted waves, we anticipate that the period of resistance oscillations in $k$ should scale inversely with $L$. Because the HWHM $x^*$ can be comparable to $L$, it is not clear what the true electronic length of the cavity is. This uncertainty is less acute for larger $L$, as we expect that $x^*\approx 200$ nm not only for the $L = 320$ nm local gate, but also for larger local gates, as the sharpness of the density profile is determined by the interplay of electric fields at the edge of the local gate, independent of $L$.

\section{Density dependence in the bulk and beneath the local gates}
The density in the regions lying exclusively below the top gate is shown as a function of $V_{\rm tg}$ in Supplementary Fig.~S2a. The total, electron, and hole densities, denoted by $n_{\rm tot} = n - p$, $n$ and $p$, respectively, are piecewise linear functions of $V_{\rm tg}$, in accordance with a capacitor model for the InAs/GaSb double QW system. At the charge neutrality point (CNP) at $V_{\rm tg} = -1.03$V, $n = p$ such that $n_{\rm tot} = 0$.  We determine $n_{\rm tot}$ from the classical Hall effect by fitting the transverse resistivity $\rho_{xy}$ at sufficiently high magnetic fields where $\rho_{xy}$ is linear in $B$ and inversely proportional to $n_{\rm tot}$. We find $n$ from the frequency of Shubnikov–de Haas (SdH) oscillations occurring in the longitudinal resistivity $\rho_{xx}$ at low magnetic fields. The dependencies shown in Supplementary Fig.~S2a are best fits to the data points obtained in this fashion. The hole density $p$ is calculated according to $p = n - n_{\rm tot}$, as it cannot be measured directly; the SdH oscillations of the holes are too poor in quality.

In order to measure $\rho_{xx} = V_{xx}/I W/L_l$ and $\rho_{xy} = V_{xy}/I$, we use voltage probes specifically designed to this end, refer to Extended Data Fig.~2.

Supplementary Fig.~S2b depicts the analogous density dependence, but below the local gate with $L = 320$nm. The CNP is located at $V_{\rm lg} = -0.72$V, and $p'$ vanishes at $V_{\rm lg} = 1.12$V. Here, we do not have direct access to $\rho_{xx}$ and $\rho_{xy}$ of the locally gated region, and the density determination is more involved. To find $n'$, we employ low-field SdH oscillations in $dR_j/dV_{\rm lg}$, which reflect the density below the locally gated region only. For $n'_{\rm tot}$, we analyze the Landau fan diagram of $R_j$ or $dR_j/dV_{\rm lg}$ as a function of $B$ and $V_{\rm lg}$ at high magnetic fields, in which the quantum Hall effect becomes visible. There, minima in $R_j$ occur at integer total filling of the locally gated region, and from their positions in ($V_{\rm lg}$, $B$) space we infer $n'_{\rm tot}$. The hole density $p'$ is again calculated according to $p' = n' - n'_{\rm tot}$. We set $V_{\rm tg} = -1.8$V in all of the above measurements.

The regions below the other local gates of device A exhibit similar density dependencies as shown in Supplementary Fig.~S2b, as expected from a homogeneous device.

\section{Reproducibility of the oscillations in a single device and length dependence}
The resistance oscillations in the pn’p configuration, which we attribute to Fabry–P{\'e}rot interference, are robust. In Supplementary Fig.~S3, we present a zoom-in of said oscillations for the cavity with $L = 320$nm. The two data sets  are captured months apart, in different cooldowns, and demonstrate the robustness and reproducibility of the interference phenomenon, as well as the quality and stability of our devices.

In Supplementary Fig.~S4 we expound the length dependence of the interference. At $1.3$K, the oscillations are also clearly visible for the gate of nominal length $L = 400$ nm, see Supplementary Fig.~S4a. Their behavior is similar to that occurring for the $L = 320$ nm gate (cf.~Fig.~1d of the main text). Namely, the oscillation amplitude decreases with increasing energy, towards more positive $V_{\rm lg}$, while the period simultaneously tends to decrease. The oscillations again disappear around $V_{\rm lg}\sim 1$V, which is roughly where $p'$ vanishes. The oscillation amplitude is in general reduced in comparison with the $L = 320$ nm case, as we will discuss shortly in a more quantitative manner (Supplementary Fig.~S4d).

Supplementary Fig.~S4b depicts the phase diagram when the final local gate of device A, with $L = 540$ nm, is energized. At $1.3$K, no reproducible oscillations are visible anywhere. Therefore, we only measured the phase diagram in a reduced range to save time.

Decreasing the temperature $T$ increases the amplitude of the oscillations. The temperature dependence is analyzed in detail in one of the next sections. In Supplementary Fig.~S4c, we show $dR_j/dV_{\rm lg}$ at $V_{\rm tg} = -1.8$V for all three local gates of device A and at $T = 60$ mK. Now, there are also weak, but reproducible oscillations visible when probing the region under the largest gate. We restrict ourselves to the interval $-0.85{\rm V} < V_{\rm lg} < 1$V and extract a mean oscillation amplitude for all three gates, and summarize the result in the semi-logarithmic plot of Supplementary Fig.~S4d. We also fit the length dependence of the mean oscillation amplitude with a function of the form $A_0 \exp(-2L/l_\varphi)$, where $A_0$ is a constant prefactor and $l_\varphi$ is the phase coherence length. We obtain $l_\varphi=424 \pm100$ nm  for the phase coherence length at $60$ mK. The large error in $l_\varphi$ reflects the uncertainty in the determination of the amplitude, as indicated by the error bars in Supplementary Fig.~S4d. This uncertainty arises due to the fact that the amplitude is not constant in the chosen voltage interval, as noted above. This actuality, together with the fact that the exact electronic length of the cavity is unknown, see Supplementary Fig.~S1 and the accompanying discussion, implies that the $l_\varphi$ reported here should be understood as an order of magnitude estimate.

To find the mean oscillation amplitude and its uncertainty (standard deviation) for a given $L$, we first placed two interpolating cubic splines through the local minima and maxima of $dR_j/dV_{\rm lg}$, respectively. Then, taking their average gave us the signal background. The amplitude is then calculated for each minimum and maximum by subtracting this background from the data. Finally, we average over all minima and maxima in the interval $-0.85{\rm V} < V_{\rm lg} < 1$ V, thereby obtaining a mean and standard deviation.

\section{Bias current dependence}
Supplementary Fig. S5 shows the bias dependence of the oscillations in $dR_j/dV_{\rm lg}$ in the $pn'p$ configuration, at $V_{\rm tg} = -1.8$ V, as a function of $V_{\rm lg}$. The oscillations are symmetric with respect to the reversal of the current direction (Supplementary Fig. S5a). Increasing the current I suppresses the interference, and the oscillation amplitude decreases in a nonlinear manner due to a combination of resistive heating and bias broadening (Supplementary Fig. S5b). The resistive heating locally increases the temperature, and is equivalent to warming up the device, as described in the next section. The bias broadening refers to the opening of the bias window, leading to averaging in energy over multiple modes, ultimately reducing the oscillation amplitude. 

Below 50 nA the amplitude saturates, but the signal to noise ratio is poor. For the majority of our experiments, we chose $I = 100$ nA as a compromise between signal quality and amplitude suppression.  We did not apply currents above 600 nA, equivalent to a current density of 120 nA $\mu$m$^{-1}$, in fear of irreversibly damaging the device. 

We cannot obtain an energy scale from the observed current dependence for various reasons. The decrease of the amplitude for the applied biases is too small. Additionally, we do not precisely know how to convert I to the bias window energy because the resistance of the cavity is unknown, and therefore we do not know the potential difference over the cavity. In fact, this resistance is a function of both $V_{\rm lg}$ and $V_{\rm tg}$.

\section{Temperature dependence}
Increasing the temperature leads to a broadening of the Fermi–Dirac distribution around the Fermi energy, such that the energetic interval relevant for transport increases. In the context of our interferometer, this phenomenon leads to averaging over multiple adjoining standing wave modes, ultimately reducing the modulation of the resistance. Supplementary Fig. S6a illustrates this suppression of the oscillation amplitude for the local gate of length $L = 320$ nm as the temperature is gradually increased from 50 mK to 8~K at a fixed top gate voltage, $V_{\rm tg} = -1.8$ V. Above 4~K the oscillations vanish, and below $\sim$ 500 mK their amplitude saturates.  

For a more quantitative analysis, we determine the oscillation amplitude as a function of temperature for the various minima and maxima marked by symbols in Supplementary Fig. S6a. We employ the same technique as described in the section on length dependence, which consists of placing splines through the minima and maxima to determine the background and subtracting this background to obtain the amplitude. Supplementary Fig. S6b shows the temperature dependence of the amplitude of the maximum located at $V_{\rm lg} = -0.365$ V, as marked by the arrow in Supplementary Fig. S6a, together with a fit. The fit is given by a thermal damping function of the form
	\begin{equation}
		A=A_0 \frac{2 \pi^2 k_B T}{\Delta E} \, \frac{1}{\sinh \frac{2 \pi^2 k_B T}{\Delta E}  }
	\end{equation}
where  $A_0$ is a constant prefactor and $\Delta E$ is the characteristic mode spacing in the cavity. We repeat the fitting procedure for the other extrema and summarize the results in Supplementary Fig. S6c. We recognize that $\Delta E$ is constant in the investigated voltage (energy) range, obtaining $\Delta E = 0.97 \pm 0.07$ meV upon taking the mean and standard deviation of all data points. This result is in agreement with theoretical analysis of the cavity mode spacing.

\section{Oscillation period analysis}
Supplementary Fig. S7a is an index plot of the oscillation extrema in $d R_j/d V_{\rm lg}$ for the local gate of length $L = 320$ nm, taken at 60 mK (trace in top panel of Supplementary Fig. S4c). We restrict ourselves to the interval $-0.7 < V_{\rm lg} < 1$ V. The index i counts the minima and maxima in order of occurrence. More concretely, the first minimum in the aforementioned interval is labeled with $i = 1$, the first maximum with $i = 2$, the second minimum with $i = 3$, and so on. We fit all points thus obtained with a linear function. The corresponding slope multiplied by two (to account for having both minima and maxima) gives the average (global) period $\Delta V_{\rm lg}$ of the oscillations in $V_{\rm lg}$. We obtain $\Delta V_{\rm lg} = 160$ mV, recognizing however that the oscillations are not truly periodic in $V_{\rm lg}$, as the points systematically deviate from the fit in places. This deviation is exemplified in Supplementary Fig. S7b, where we calculate the local period by taking the difference between subsequent minima (circles) and subsequent maxima (diamonds) from (A), noting that $\Delta V_{\rm lg}$ is not a monotonic function of $V_{\rm lg}$. We superimpose the expected dependencies if the oscillations were exactly periodic in $k_n - k_p$ (dash-dotted line) or $2 k_n$ (dashed line). Varying $L$ simply shifts the curves up and down but does not change their shape. By comparing the evolution of the experimentally determined $\Delta V_{\rm lg}$ and the expected $\Delta V_{\rm lg}$, we conclude that the oscillations are not periodic on either axis. This is not completely unexpected, and signals that the interference arises due to a combination of competing processes involving electron-hole and electron-electron/hole-hole scattering, as elucidated in the main text. 

We now convert the gate voltage axis to new axes $k_n - k_p$ (Supplementary Fig. S7c) as well as $2 k_n$ (Supplementary Fig. S7d) and study the modified index plots. Motivated by discussions presented in the main text, we determine the period on the $k_n - k_p$ axis at low energies, close to the CNP. To this end, we fit the first six points of the index plot in Supplementary Fig. S7c with a linear function. As before, the slope gives the average period $\Delta(k_n - k_p)$. Then, $2 \pi/\Delta(k_n - k_p) = 490$ nm should equal the length $L$ of the gate. In light of the uncertainty in the effective gate length due to the smooth electrostatic profile, which is described by the HWHM $x^* = 200$ nm (see above), this value agrees well with the geometric gate length of $L = 320$ nm, or more precisely, with $L \pm x^* = 320 \pm 200$ nm.

We repeat the same analysis on the $2 k_n$ axis at high energies where the holes are almost depleted  by fitting the last six points of the index plot in Supplementary Fig. S7d. We get $2 \pi/\Delta(2 k_n) = 400$ nm, which is once again consistent with expectation.

Lastly, we duplicate the above analysis for the other local gates of device A. The results agree with those obtained for the $L = 320$ nm gate. The average period $\Delta V_{\rm lg}$ decreases from 160 to 140 to 100 mV as $L$ increases from 320 nm to 400 nm to 540 nm. We obtain $2 \pi/\Delta(k_n - k_p) = 560$ nm and $2 \pi/\Delta(2 k_n) = 450$ nm compared to $L \pm x^* =  400 \pm 200$ nm. Likewise, we have $2 \pi/\Delta(k_n - k_p) = 680$ nm when $L \pm x^* =540 \pm 200$ nm, opting not to provide an estimate of the period in $2 k_n$ as the determination of the minima and maxima at high energies is ambiguous due to poor signal to noise ratio for the local gate of length $L = 540$ nm.

\section{Magnetic field dependence}
Increasing the perpendicular magnetic field $B$ eventually leads to Landau quantization in the cavity. Supplementary Fig. S8 shows the numerical derivative of the measured quantity $d R_j/d V_{\rm lg}$ with respect to $V_{\rm lg}$ as a function of $V_{\rm lg}$ and $B$, where we deliberately employ a saturated color scale to enhance the visibility. With increasing B two types of regular features appear. The first, marked by the arrows, are lines corresponding to SdH oscillations of the electron-like states in the cavity. The slopes of the lines agree with the independently determined electron density (Supplementary Fig. S2). The second, indicated by the solid lines, are associated with the emergence of the quantum Hall state in the cavity.  Indeed, minima in the resistance $R_j$ appear at integer total filling factor $\nu = \nu_e - \nu_h$ where $\nu_e$ and $\nu_h$ are the filling factors of electron-like and hole-like Landau levels, respectively. The minima lie on a background which increases with $B$ and therefore by no means reach zero resistance. We speculate that the quantum Hall state in the cavity is not well-defined due to its size, and that the quantum Hall edge states in the neighboring leads leak through it. 

The lines associated with the quantum Hall state in the cavity connect to the zero-field Fabry–Pérot oscillations. Apart from a transition region in which the lines may become interrupted, we may track the resistance oscillations starting from $B = 0$ all the way up to the highest magnetic fields (solid lines in Supplementary Fig. S8). In graphene, this phenomenon has been interpreted in terms of disorder-mediated SdH oscillations \cite{young_quantum_2009}. 

\section{Reproducibility of the oscillations across multiple devices}
Supplementary Fig. S9 showcases the interference phenomenon in device B. As explained above, device B has the same QW thicknesses as device A, which is the primary device investigated. However, device B comprises only two working local gates with $L = 530$ nm and $L = 700$ nm. Reproducible resistance oscillations emerge when probing both gates. In agreement with our experience on device A, the oscillations are weak at 1.3 K for these gate lengths. Therefore, in Supplementary Fig. S9 we present color maps of $d R_j/d V_{\rm lg}$ as a function of $V_{\rm lg}$ and $V_{\rm tg}$ recorded at 60 mK. In the $L = 530$ nm case (Supplementary Fig. S9a), regular oscillations are clearly visible in the $pn'p$ configuration. That is not so in the $np'n$ configuration, as reported for device A, despite the fact that here the gate voltage range in $V_{\rm lg}$ extends deeper into the hole regime than for device A. 

In the $L = 700$ nm case (Supplementary Fig. S9b), reproducible oscillations are also visible in the same gate voltage space, but their amplitude is reduced. 

Overall, the resistance oscillations seen in device B exhibit the same behavior as reported in device A. This consistency demonstrates the robustness of the interference phenomenon in our hybridized electron-hole bilayers. 

We have also discovered resistance oscillations due to quantum interference in device C, which has a heterostructure featuring a thinner InAs QW, leading to a shallower band inversion and therefore to a lesser degree of hybridization. The strength of the oscillations is lower than in devices A, B for the same or comparable $L$ and at the same temperature, as summarized in Supplementary Fig. S10. Clear, reproducible oscillations are present when probing the gate of length $L = 320$ nm at 1.3 K at low densities close to the CNP of the cavity (Supplementary Figs. S10a, b). With increasing voltage $V_{\rm lg}$ the oscillations vanish quickly; note the reduced range in $V_{\rm lg}$ compared to device A. Despite the poorer visibility, the measurements are consistent with those on devices A, B. The oscillatory features appear in the $pn'p$ configuration and they have a slightly negative slope in the ($V_{\rm lg}$, $V_{\rm tg}$) plane, the oscillation period increases and the amplitude decreases with increasing $V_{\rm lg}$ and the amplitude depends on the length $L$ of the local gate. The last point is exemplified in Supplementary Figs. S10c and d and Supplementary Figs. S10e and f for the local gates of length $L = 400$ nm and $L = 540$ nm, respectively. In the former case, we are still able to make out some regular oscillations, whereas in the latter, this becomes impossible.

There exist multiple explanations as to why the interference effect is weaker in device C compared to devices A, B. The most trivial reason is simply coincidence, i.e., naturally occurring variations in device quality. Another reason involves differences in mobility, and therefore perhaps phase coherence length. In our experience, the heterostructure with $d_{\rm InAs} = 13.5$ nm typically has average Drude mobilities which are roughly two to three times greater than those in the heterostructure with $d_{\rm InAs} = 12.5$ nm at comparable densities in the electron dominated regime. Lastly, the hybridization of electrons and holes is weaker when the InAs QW thickness is reduced as this decreases the band overlap. As we conjecture that the hybridization plays a key role in facilitating the interference, we would indeed expect to see a difference in visibility in this case.

\end{document}